\documentclass[preprint,12pt,authoryear]{elsarticle}

\usepackage{hyperref}
\usepackage{aas_macros}
\usepackage{textcomp}
\usepackage{amsmath}

\journal{Icarus}

\biboptions{authoryear}

\DeclareSymbolFont{symbolsC}{U}{txsyc}{m}{n}
\SetSymbolFont{symbolsC}{bold}{U}{txsyc}{bx}{n}
\DeclareFontSubstitution{U}{txsyc}{m}{n}
\DeclareMathSymbol{\nll}{\mathrel}{symbolsC}{51}

\begin{document}

\begin{frontmatter}

\title{Escape and fractionation of volatiles and noble gases from Mars-sized planetary embryos and growing protoplanets}

\author[iwf]{P.~Odert\corref{cor1}}
\ead{petra.odert@oeaw.ac.at}

\author[iwf]{H.~Lammer}

\author[icm,sfu]{N.~V.~Erkaev}

\author[dlr,tub]{A.~Nikolaou}

\author[iwf]{H.~I.~M.~Lichtenegger}

\author[uv]{C.~P.~Johnstone}

\author[uv,iwf]{K.~G.~Kislyakova}

\author[igam]{M.~Leitzinger}

\author[dlr,tub]{N.~Tosi}

\cortext[cor1]{Corresponding author}

\address[iwf]{Space Research Institute, Austrian Academy of Sciences, Schmiedlstrasse 6, 8042 Graz, Austria}

\address[icm]{Institute of Computational Modelling of the Siberian Branch of the Russian Academy of Sciences, 660036 Krasnoyarsk, Russian Federation}

\address[sfu]{Siberian Federal University, 660041 Krasnoyarsk, Russian Federation}

\address[dlr]{Institute for Planetary Research, German Aerospace Center (DLR), Rutherfordstrasse 2, 12489 Berlin, Germany}

\address[tub]{Technical University Berlin, Strasse des 17. Juni 135, 10623 Berlin, Germany}

\address[uv]{University of Vienna, Department of Astrophysics, T\"urkenschanzstrasse 17, 1180 Vienna, Austria}

\address[igam]{Institute of Physics, IGAM, University of Graz, Universit\"atsplatz 5, 8010 Graz, Austria}

\begin{abstract}
Planetary embryos form protoplanets via mutual collisions, which can lead to the development of magma oceans. During their solidification, significant amounts of the mantles' volatile contents may be outgassed. The resulting H$_2$O/CO$_2$ dominated steam atmospheres may be lost efficiently via hydrodynamic escape due to the low gravity of these Moon- to Mars-sized objects and the high stellar EUV luminosities of the young host stars. Protoplanets forming from such degassed building blocks after nebula dissipation could therefore be drier than previously expected. We model the outgassing and subsequent hydrodynamic escape of steam atmospheres from such embryos. The efficient outflow of H drags along heavier species like O, CO$_2$, and noble gases. The full range of possible EUV evolution tracks of a young solar-mass star is taken into account to investigate the atmospheric escape from Mars-sized planetary embryos at different orbital distances. The estimated envelopes are typically lost within a few to a few tens of Myr.

Furthermore, we study the influence on protoplanetary evolution, exemplified by Venus. In particular, we investigate different early evolution scenarios and constrain realistic cases by comparing modeled noble gas isotope ratios with present observations. Isotope ratios of Ne and Ar can be reproduced, starting from solar values, under hydrodynamic escape conditions. Solutions can be found for different solar EUV histories, as well as assumptions about the initial atmosphere, assuming either a pure steam atmosphere or a mixture with accreted hydrogen from the protoplanetary nebula. Our results generally favor an early accretion scenario with a small amount of residual hydrogen from the protoplanetary nebula and a low-activity Sun, because in other cases too much CO$_2$ is lost during evolution, which is inconsistent with Venus' present atmosphere. Important issues are likely the time at which the initial steam atmosphere is outgassed and/or the amount of CO$_2$ which may still be delivered at later evolutionary stages. A late accretion scenario can only reproduce present isotope ratios for a highly active young Sun, but then unrealistically massive steam atmospheres (few kbar) would be required.
\end{abstract}


\end{frontmatter}

\section{Introduction}
In the early evolution of planetary systems, protoplanetary cores originate from the coagulation of dust and ice and initially reside embedded in the gas of the circumstellar disks. Terrestrial planet-formation models indicate that Earth-like planets originate from differentiated planetesimals to large planetary embryos with sizes of several hundred to a few thousand kilometers \citep[e.g.][]{Kokubo00, Raymond04, Raymond09, Alibert10, Lunine11, Walsh11, Morbidelli12}. Although the processes that are responsible for the growth of the solid bodies from centimeter size up to the size of Moon- and Mars-size planetary embryos are not well understood \citep{Morbidelli09, Johansen14}, if one accepts the scenario of core-accretion for the formation of gaseous Jovian-type planets \citep{Perri74, Mizuno80}, massive planetary embryos and even protoplanetary cores with several Earth-masses ($M_{\oplus}$) exist even at an early stage of evolution of the protoplanetary disk. The initial volatile inventory, including H$_2$O, of terrestrial planets is related to a complex interplay between:
\begin{itemize}
\item the nebula dissipation time,
\item the growth rate/time from planetesimals and planetary embryos to protoplanets,
\item the orbit location and H$_2$O content of the initial building blocks (i.e., planetesimals and planetary embryos),
\item outgassing processes from the interior,
\item the impact history,
and
\item the host star's radiation and plasma environment.
\end{itemize}
The growth of planetary embryos to protoplanets within the accretion disk begins during the nebula lifetime, which lasts only between $\approx$1--10~Myr \citep{Montmerle06, Hillenbrand08}. Mars for instance formed within a few Myr and remained as a large planetary embryo that never grew to a more massive rocky planet \citep{Brasser13}. Depending on the gravitational potential of the embedded planetary embryos and protoplanets, a certain amount of the nebula gas will be captured. As it was shown by \citet{Erkaev14} and \citet{Stoekl15}, cores with masses up to $\approx 0.1M_{\oplus}$ capture only a negligible amount of nebula gas which is lost fast by thermal escape shortly after the nebula dissipates. Depending on the solar nebula parameters and the EUV activity of the young Sun, \citet{Erkaev14} showed that early Mars which has a mass of $\approx 0.1M_{\oplus}$ should have lost its captured nebular-based hydrogen envelope in a very short time. For less massive bodies, or objects at closer orbital distances, the timescales are even shorter because of either the lower gravitational attraction or the higher EUV fluxes and effective temperatures. Thus, for planetary embryos, accumulated nebular gas does not play an important role for the bodies' evolution into larger planetary objects.

After the evaporation of the gas in the disk due to the extremely high X-ray, EUV and far ultraviolet (FUV) emissions of the young T~Tauri Sun/star, protoplanets continue to grow through the capture and collisions of large planetesimals and planetary embryos. Depending on the orbit locations, embryos may form dry or wet, meaning that near and beyond the ice line the planetary building blocks contain more water and icy material compared to that which originated and orbit closer to the Sun. From dynamical models one can expect that most of Mars' building blocks consisted of material that formed in orbital locations just beyond the ice line which could have contained $\approx 0.1-0.2$~wt\% of H$_2$O, while embryos formed in Earth's orbit should have had lower values within a range of $\approx 0.05-0.1$~wt\% \citep[e.g.][]{Morbidelli00, Brasser13}. In the case of the Earth, geochemical studies indicate that a fraction of Earth's initial H$_2$O inventory originated from comets, while the majority came from chondritic meteoritic materials \citep{Mumma11, Alexander12, Marty12}. Meteorites are therefore the main candidates for building blocks of planetary embryos and hence protoplanets.

The study of different solar system objects that resemble planetary embryos (such as Earth's Moon, Mars, the asteroid Vesta, and dwarf planets like Ceres) shows that they have been differentiated in their interiors like the terrestrial planets \citep[e.g.][]{Canup02, Schubert04, Thomas05, Russell13}. It is thus reasonable to expect that the silicate and metal materials of large planetary embryos of similar size were largely molten during the formation process \citep[e.g.][]{Albarede07, Elkins-Tanton08a, Elkins-Tanton12}.

The formation of large and deep magma oceans can also be caused by gravitational heating of the accreted material, accretionary impacts, and radiogenic heating from short-lived radioisotopes, \citep[e.g.][]{Urey55, Safronov69, Wetherill80, LaTourrette98, Halliday01, Albarede07, Elkins-Tanton12}. Magma oceans are responsible for the compositional differentiation that affects the final volatile contents of the planetary building blocks, including large planetesimals, planetary embryos, protoplanets or large moons, whose radii range from tens to hundreds, or even thousands of kilometers. The melting related to the differentiation results in compositionally distinct layers, where the denser materials sink to the center, and less dense materials rise to the surface. Such processes finally create a core and mantle in planetary bodies.

Volatiles such as H$_2$O and carbon compounds (e.g., CO$_2$, CH$_4$, CO) are integrated in the magma ocean liquids and as solidification proceeds they are degassed into a growing steam atmosphere \citep{Matsui86, Abe88, Zahnle88, Massol16}. The quantity of volatiles available for degassing depends on the bulk composition of the magma ocean. The solidification of a magma ocean starts at the bottom because the steep slope of the adiabat with respect to the solidus in the pressure-temperature space causes them to intersect first at depth \citep[e.g.][]{Walker75, Solomatov00, Elkins-Tanton08a, Elkins-Tanton11, Elkins-Tanton12}. Depending on the assumed H$_2$O ($\approx 0.05{-}0.1$~wt\%) and CO$_2$ (${\approx}0.01{-}0.02$~wt\%) content with bulk magma ocean depths between ${\approx}500{-}2000$~km for a Mars-size body, minimum and maximum partial surface pressures $P_\mathrm{H_{2}O}$ and $P_\mathrm{CO_{2}}$ of catastrophically outgassed steam atmospheres are estimated to be ${\approx}30{-}120$~bar and ${\approx}7{-}25$~bar \citep{Elkins-Tanton08a, Lebrun13, Erkaev14}.

As pointed out by \citet{Lammer13a}, depending on the EUV flux evolution of the young Sun/star, Moon- or Mars-sized planetary embryos will lose a fraction of their outgassed volatiles and their initial water inventories because of the formation of magma oceans within the growth process to larger protoplanets. It has been known for decades that the Sun's radiation in the EUV region of the spectrum was higher in the past \citep[e.g.][]{Zahnle82, Guedel97, Guinan02, Ribas05, Claire12}. Recently, \citet{Tu15} showed that the star's initial rotation rate and its subsequent rotational evolution \citep{Johnstone15} play an important role for the EUV flux enhancement, until the time when the radiation flux converges to similar age dependent values after about 1.5~Gyr in the case of solar-like G~stars. About 70\% of the solar mass stars studied by \citet{Johnstone15} are slow and moderate rotators, but there is a non-negligible possibility for the early Sun to have been a fast rotator. These findings have enormous implications for the initial volatile inventories of terrestrial planets and, later on, their atmospheric evolution.

If one assumes that the young Sun was initially a fast rotator, then its EUV emission could have been enhanced up to about 500 times compared to its present value for more than 200~Myr. If the Sun was once a moderate rotator then the EUV emission would have been about 100~times higher during the first 100~Myr. If the Sun was a slow rotator then its EUV luminosity would have been about 25--30~times higher for $\approx 30-300$~Myr. According to \citet{Tu15}, the EUV flux decreases for all rotators following power laws until the different evolutionary paths merge after about 1.5~Gyr. Because the EUV radiation heats the upper atmospheres of planets one can expect a huge variety of evolution scenarios depending on the stars' different rotation and activity evolution during the first 1.5~Gyr. As a consequence, planetary embryos which form and grow in a system where the host star is a slow rotator should remain wetter and more volatile-rich compared to those which grow in a system of moderately or fast rotating young stars.

The discovery of a large number of sub-Neptune type exoplanets indicates that fast formation of sufficiently large cores which then accrete residual nebula gas is common. \citet{Stoekl16} found that a core of $0.7M_\mathrm{Venus}$ ($\sim0.6M_{\oplus}$) may accrete gas of up to 2--3\% of its mass within a typical nebula lifetime. The core luminosity can drive efficient outflow of such accreted gas envelopes when the confining pressure of the disk decreases, but cores with masses $>0.5M_{\oplus}$ are rather stable to this loss process \citep{Stoekl15, Stoekl16}. After the disk dissipates at the orbital location of the protoplanet a significant fraction of such a protoatmosphere can be lost in timescales of a few Myr by ``boil-off'' driven by the envelope contraction and bolometric luminosity of the host star \citep{Owen16, Ginzburg16, Lammer16} and enhanced by the initially very extended envelopes (comparable to the Bondi radius, i.e. $r_0\sim GM_p/(2c_s^2)$). Due to this efficient mass-loss, the protoplanet's H atmosphere quickly shrinks until reaching a more stable escape regime. Recent studies found that this occurs when the radius shrinks below 0.1 Bondi radii \citep{Owen16} or, equivalently, when the Jeans escape parameter at $r_0$ becomes larger than about 20 \citep{Fossati17, Cubillos17}.

The recent Ru isotope analysis of primitive meteoritic material \citep{Fischer-Goedde17} indicates that volatile-rich carbonaceous chondritic bodies were scattered from the outer into the inner solar system early in Earth's, Venus' and Mars' accretion phase, contradicting previous studies of \citet{Albarede09}. Therefore, one can expect that these bodies originated in a very cold environment beyond the ice-line and remained most likely volatile-rich. This scenario is in agreement with dynamical model results related to the ``Grand Tack'' hypothesis \citep{Walsh11, OBrien14}. In this model, Jupiter migrated inwards to the terrestrial planet region and subsequently outwards to its present location due to the influence of Saturn. This resulted in scattering of material to the terrestrial planet region, but cleared the disk down to about 1~AU, which can explain why Mars remained small \citep{Walsh11}. If this migration scenario of Jupiter scattered volatile-rich primitive planetesimals towards the inner planets, numerous large volatile-rich carbonaceous chondritic impactors may have collided with the early terrestrial proto-planets. In such a case their volatiles could have been delivered efficiently to proto-Venus, proto-Earth and early Mars \citep{OBrien14}.

Regardless of the detailed EUV history of the host star, embryos generally lose their potential atmospheres faster than more massive bodies because of their lower gravities. Moreover, embryos in orbits closer to the star are more strongly affected because the stellar EUV fluxes are higher. This means that if close-in planets such as Venus accreted from embryos which have been located near the planet's orbit during a few million years, they may have accreted from drier material. Large impacts occurring after several tens of Myr could possibly only deliver negligible amounts of volatiles to the growing planets at such orbital distances, except if they have just recently been scattered inwards from the outer regions of the system.

Another consequence of efficient atmospheric escape on planetary embryos or protoplanets is the fractionation of noble gases \citep[e.g.][]{Pepin91}. Both element abundances and isotope ratios of a single element are expected to be altered due to the escape of a protoatmosphere. \citet{Donahue86} investigated the escape from planetesimals and related isotope fractionation via Jeans escape. However, this author suggested, that while being possible, it would be more likely that a less mass-dependent process would be better suited. This was also noted by \citet{Hunten87} who favored  hydrodynamic escape due to its weaker mass-dependence and because it was able to produce results that are comparable with observations on various planets. Numerous studies aimed to explain the observed noble gas patterns on the terrestrial planets via escape of a protoatmosphere during the first evolutionary stages \citep[e.g.][]{Sekiya80, Hunten87, Sasaki88, Zahnle90, Pepin91, Gillmann09}. However, the possibility that the building blocks forming these planets could have been fractionated before by hydrodynamic escape has not been studied yet in detail.

This work is the first attempt to address these important aspects that have received so far little attention. We estimate the escape of H$_2$O and CO$_2$ atmospheres resulting from magma ocean degassing of Mars-sized planetary embryos at orbital locations of Venus (0.7~AU), Earth (1~AU) and Mars (1.5~AU) that are exposed to the EUV emission of a slowly, moderately and fast rotating young solar-like star. In Section~\ref{sec:model} we describe the applied model and input parameters. Subsequently, we present the results, the atmospheric evolution of Mars-like embryos (Section~\ref{sec:embryos}) and a proto-Venus scenario including constraints from noble gas isotope fractionation (Section~\ref{sec:venus}). In Section~\ref{sec:disc} we discuss the results and model limitations, and present the conclusions in Section~\ref{sec:conclusions}.

\section{Model description}\label{sec:model}

Stellar EUV photons within the wavelength range of $\sim$10--91.2~nm are the main heating source in the upper atmospheres of planetary bodies \citep[e.g.][]{Watson81, Chassefiere96a, Chassefiere96, Tian09, Tian15a, Erkaev14, Erkaev16}. Stellar FUV emission, including the Schumann-Runge continuum, is another important heating source dominating in the lower thermosphere of present Earth via dissociation of O$_2$ \citep{Roble87}. However, it is not clear how efficient this process is in a massive hot steam atmosphere since no detailed photochemical models are available for such atmospheres. Moreover, stellar FUV is enhanced only by factors of a few in young stars compared to today's value, whereas X-rays and EUV can be higher by factors of a few tens or hundreds \citep{Claire12, Tu15}. Therefore, we ignore this potential additional heating source in our present model. After the phase of catastrophic volatile outgassing in the form of H$_2$O and CO$_2$ from planetary embryos during their magma ocean solidification phase, the upper atmosphere is heated due to absorption of the incoming EUV radiation, leading to excitation, dissociation and ionization of the gas components. The high EUV fluxes of the young star will therefore lead to the dissociation of H$_2$O and CO$_2$ molecules \citep[e.g.][]{Chassefiere96, Tian09, Lammer13b}. Because of the efficient dissociation of H$_2$O one can expect that planetary embryos have initially a hydrogen-dominated upper atmosphere. Such an atmosphere will expand hydrodynamically and escape efficiently \citep[e.g.][]{Watson81, Chassefiere96, Chassefiere96a, Tian05, Lammer13a, Lammer14, Erkaev13, Erkaev14}.

\subsection{Stellar EUV emission}
\label{sec:euv}
Stellar EUV emission is the main driver of atmospheric escape in young planetary systems. A detailed study of the evolution of the stellar EUV output of Sun-like stars was performed by \citet{Tu15}. For stars younger than about 2~Gyr, the intrinsic spread in stellar rotation rates and corresponding high-energy emission is large. To explore all possible scenarios we model atmospheric escape for three cases of stellar EUV evolution tracks, corresponding to the 10$^\mathrm{th}$, 50$^\mathrm{th}$, and 90$^\mathrm{th}$ percentiles of the rotation rate distribution (hereafter termed slow, moderate, and fast rotators) from \citet{Tu15}. Stars spin down with age because of angular momentum loss via magnetized stellar winds. Since rotation and magnetic activity (e.g. in form of the high-energy X-ray and EUV emission) are closely related in Sun-like stars via the magnetic dynamo, this leads to a decrease in magnetic activity with rotation rate over time. At young ages, however, the stellar X-ray and EUV emission is saturated at a roughly constant fraction of the star's bolometric luminosity ($L_X\sim10^{-3}L_\mathrm{bol}$). This saturation phase lasts for about 5.7, 23, and 226~Myr for the slow, moderate, and fast rotator tracks. Afterwards the EUV luminosity $L_\mathrm{EUV}$ (100--920~\AA) decays as
\begin{equation}\label{eq:euv}
	L_\mathrm{EUV}=
	\begin{cases}
		7.4\times10^{31}t^{-0.96}\hspace{2cm}10^\mathrm{th}\\
		4.8\times10^{32}t^{-1.22}\hspace{2cm}50^\mathrm{th}\\
		1.2\times10^{36}t^{-2.15}\hspace{2cm}90^\mathrm{th},
	\end{cases}
\end{equation}
where $t$ is the stellar age in Myr. The saturation value $L_\mathrm{EUV,sat}$ can be found from Eq.~\ref{eq:euv} by inserting the corresponding saturation times, which yields a common value of $\sim10^{31}$~erg~s$^{-1}$ for a solar-like star. Hereafter we will refer to these minimum, average and maximum EUV evolution tracks as the slow, moderate and fast rotator scenarios.

\subsection{Escape fluxes and hydrodynamic drag}\label{sec:escape}
Under the high EUV fluxes of young stars, atmospheres of planetary embryos and protoplanets escape hydrodynamically \citep{Watson81, Kasting83, Erkaev14, Erkaev16}. A simple estimate of the hydrodynamic escape rate is given by the energy-limited equation \citep{Watson81, Erkaev07}. For a three-component atmosphere, it can be written as
\begin{equation}\label{eq:mdot}
m_{i}F_{i} + m_{j}F_{j} + m_{k}F_{k} = \frac{\beta^2 \eta F_\mathrm{EUV}}{4\Delta\Phi} \equiv \dot{m},
\end{equation}
where $m_{i,j,k}$ are the particle masses, $F_{i,j,k}=n_{i,j,k}v_{i,j,k}$ the escape fluxes at $r_0$ (in cm$^{-2}$~s$^{-1}$~sterad$^{-1}$), $\eta$ is the heating efficiency, $F_\mathrm{EUV}$ the stellar EUV flux at the orbit, $\Delta\Phi=GM/r_0$ the object's gravitational potential at $r_0$, and $\dot{m}$ the total mass-loss rate (in g~cm$^{-2}$~s$^{-1}$~sterad$^{-1}$). The parameter $\beta=r_\mathrm{EUV}/r_0$ is the ratio of the effective radius $r_\mathrm{EUV}$ where the bulk of the incoming EUV radiation is absorbed to the atmospheric radius $r_0$.

We aim to study the hydrodynamic escape of atomic H which drags along heavier constituents of the upper atmosphere (e.g. O, C, CO$_2$, noble gases). Specifically, we consider one heavy major species (such as O in a dissociated H$_2$O atmosphere), as well as an arbitrary number of heavy minor species. Hereafter, we will refer to the light major species with index $i$, the heavy major species with index $j$, and the heavy minor species with index $k$. To obtain the escape fluxes of the heavier constituents, we follow the method of \citet{Zahnle86} and \citet{Zahnle90}. They write the escape fluxes of the dragged heavy species $F_{j,k}$ as
\begin{equation}\label{eq:esc}
F_{j,k}=F_{i}f_{j,k}x_{j,k},
\end{equation}
where $F_{i}$ is the escape flux of the light major species (here, H), $f_{j,k}=n_{j,k}/n_{i}\sim N_{j,k}/N_{i}$ is the mixing ratio relative to species $i$ ($n$ are the local number densities, $N$ the atmospheric inventories) and $x_{j,k}$ the fractionation factors. From comparison of Eq.~\ref{eq:esc} with the definition of the escape flux one can see that the fractionation factors correspond to the velocity ratios $v_{j,k}/v_{i}$. For $x_{j,k}$, \citet{Zahnle86} and \citet{Zahnle90} obtained analytic approximations by introducing some simplifications into the multi-species hydrodynamic equations (e.g. subsonic flow, isothermal conditions). \citet{Zahnle90} showed that these expressions provide good approximations also for transonic escape, because the relative fluxes of the species are already determined in the subsonic region.

The fractionation factors $x_{j}$ and $x_{k}$ of the heavier species can be written as
\begin{equation}\label{eq:xj}
x_{j} = 1 - \frac{g\left(m_{j}-m_{i}\right)b_{i,j}}{F_{i}k_{\rm B}T\left(1+f_{j}\right)},
\end{equation}
and
\begin{equation}\label{eq:xk}
x_{k} = \frac{1 - \frac{g\left(m_{k}-m_{i}\right)b_{i,k}}{F_{i}k_{\rm B}T} + \frac{b_{i,k}}{b_{i,j}}f_{j}\left(1-x_{j}\right) +  \frac{b_{i,k}}{b_{j,k}}f_{j}x_{j}}{1+\frac{b_{i,k}}{b_{j,k}}f_{j}},
\end{equation}
where $g$ is the gravitational acceleration at the base of the flow $r_0$, $k_{\rm B}$ is the Boltzmann constant, and $T$ the upper atmosphere temperature \citep{Zahnle90}. The expression for $x_k$ is more complicated, because the minor species feel the drag of both major constituents, but they themselves cannot influence the flow of the major gas components (or other minor elements) due to their small abundances. The binary diffusion parameters are given by $b_{\rm H,O}=4.8\times 10^{17}T^{0.75}$~cm$^{-1}$~s$^{-1}$, $b_{\rm H,CO_2}=8.4\times 10^{17}T^{0.6}$~cm$^{-1}$~s$^{-1}$ and $b_{\rm O,CO_2}=7.86\times 10^{16}T^{0.776}$~cm$^{-1}$~s$^{-1}$\citep{Zahnle86}. Equation~\ref{eq:xj} describes the fractionation of the major heavy species (here, O) and is similar to the frequently used expression derived by \citet{Hunten87}, except that it does not require that the heavy component is a minor species (i.e. $f_j{\nll}1$), and is therefore more suitable for calculating the escape of a dissociated H$_2$O atmosphere. Equation~\ref{eq:xk} describes the fractionation of the additional heavy minor species (e.g. C, CO$_2$, noble gases), which feel the drag from both major gases. If there is no heavy major species in the atmosphere, the escape of minor heavy species would be described by Eq.~\ref{eq:xj}.

The fractionation factors take values between one (all species escape efficiently with similar velocities) and zero (the heavy species cannot escape). If the latter case occurs for the major heavy species, escape of the light species is then limited by diffusion through the static, heavy background gas, and its escape flux can be found from Eq.~\ref{eq:xj} by setting $x_j=0$. If $x_k=0$, the light gas still escapes hydrodynamically, but the concerned minor species is maintained. The escape fluxes are thus given by Eq.~\ref{eq:esc} if $x_{j,k}>0$, whereas they are zero otherwise.

One can see from Eq.~\ref{eq:mdot} that the escape fluxes of all three species are connected. Combining Eq.~\ref{eq:mdot} with Eq.~\ref{eq:esc} allows to express the escape flux of the light main species as
\begin{equation}\label{eq:fi}
F_{i} = \frac{\beta^2 \eta F_\mathrm{EUV}}{4\Delta\Phi\left(m_{i} + m_{j}f_{j}x_{j} + m_{k}f_{k}x_{k}\right)},
\end{equation}
which indicates that $F_{i}$ is reduced by the presence of the heavier species compared to a single-component atmosphere. The evolution of the atmosphere is calculated by integrating the escape fluxes (Eqs.~\ref{eq:esc} and \ref{eq:fi}) in time. The stellar EUV flux (Eq.~\ref{eq:euv}), as well as the mixing ratios $f$ and fractionation factors $x$ are hereby functions of time. All other parameters are assumed to be constant.

If species $i$, which is at first the lightest major gas component, is significantly depleted from an atmosphere and is not able to drag along the other constituents anymore, then a heavier main species could under certain conditions also escape hydrodynamically. Then its escape flux could also drag along the remaining minor heavy species if the outflow is strong enough, which can be described using $i\rightarrow j$, $j\rightarrow k$, and $F_k=0$ in Eq.~\ref{eq:mdot}. We explore this scenario in more detail in Appendix~\ref{sec:oescape} using a hydrodynamic model applied to an O-dominated atmosphere.

\section{Atmospheric loss from Mars-sized embryos}\label{sec:embryos}
We investigate the loss of magma ocean degassed H$_2$O/CO$_2$ atmospheres from Mars-like planetary embryos at the orbits of Venus, Earth and Mars, exposed to the EUV emission of a slowly, moderately and rapidly rotating young Sun-like star under hydrodynamic escape conditions. We apply mass fractionation equations \citep[e.g.][]{Zahnle86, Lichtenegger16} which have first been developed in simpler form by \citet{Hunten73} and \citet{Hunten87}. Such mass fractionation scenarios have been studied in the past mainly for the investigation of the possible build-up of O$_2$ in planetary atmospheres that experience strong hydrodynamic escape of dissociation products of water \citep[e.g.][]{Chassefiere96, Lammer11, Luger15, Tian15, Tian15a}.

\subsection{Magma ocean outgassing}
\label{sec:atm}
The thermal evolution of a convecting magma ocean is studied using a one-dimensional model based on the approach of \citet{Elkins-Tanton08a} and \citet{Lebrun13}. Here we only highlight the modeling aspects relevant to the outgassing process, and refer the reader to the above papers for more details. We assume the magma ocean to solidify from its base upwards because of the steeper slope of the liquid adiabat with respect to the mantle liquidus and solidus chosen to be those of KLB-1 peridotite \citep{Zhang94, Herzberg00}. Two volatile species, water (H$_2$O) and carbon dioxide (CO$_2$), are assumed to be outgassed during the magma ocean evolution. In order for a volatile to be released into the atmosphere, its concentration in the magma ocean has to be supersaturated. Supersaturation means that the concentration of a volatile (here H$_2$O, CO$_2$) in a silicate melt should be higher than its solubility concentration. We use the saturation curves of \citet{Carroll94} for H$_2$O and \citet{Pan91} for CO$_2$, and compare the evolving concentration of the volatile in the magma ocean against them at each time step. When the magma ocean is supersaturated, the excess volatile content is outgassed in the atmosphere, building up its mass. We consider two different scenarios for the depth extent of the magma ocean: a partial magma ocean with a depth of 500 km and a global one extending to 1700~km, i.e. to the expected core-mantle boundary depth of Mars \citep{Sohl97}.

The 1D model used to estimate the volatile outgassing resolves the solidification process of an existing magma ocean and not the process of how it is generated. Therefore, the magma ocean depth and its volatile concentration are starting conditions of the simulation and have to be chosen as model parameters. The initial depth of the magma ocean is difficult to constrain due to the unknown impact history of the hypothetical embryo. By assuming global (1700~km) and partial (500~km) magma ocean depths we cover two end-member cases. The model assumes a Mars-like interior structure with a core radius of 1700~km. The final outgassed quantity depends on the initial volatile abundance of the building blocks of the planet. Bulk mantle values were chosen in order to span a broad range defined between primitive and present-day conditions: from 100 to 10000~ppm for H$_2$O, and from 10 to 1000~ppm for CO$_2$. For reference, estimates of the water concentration of the present-day Earth's upper mantle range from 50 to 200~ppm \citep{Saal02}, of the primitive mantle from 550 to 1900~ppm \citep{Jambon90}, while CI carbonaceous chondrites can contain up to 25~wt\% H$_2$O \citep{Garenne14}. The CO$_2$ content of the Earth's mantle is expected to vary between about 20 and 1300~ppm \citep{Dasgupta10}.

The cumulative abundances of outgassed H$_2$O and CO$_2$ for the various scenarios that we tested are listed in Table~\ref{tab:atm}. The partial pressures that we obtained agree with those given by \citet{Erkaev14} and \citet{Elkins-Tanton08a}, within $\approx10$\% for H$_2$O and less than 1\% for CO$_2$, because of the use of slightly different saturation curves.

\begin{table}
	\caption{Outgassed volatile budgets that result from a magma ocean stage that a Mars-sized body could have experienced.}\label{tab:atm}
	\centering
	\begin{tabular}{cccc}
		\hline
		\multicolumn{4}{c}{Partial magma ocean (500 km)} \\
		\hline
		mantle [H$_2$O]$_0$ & mantle [CO$_2$]$_0$ & outgassed $P_\mathrm{H_2O}$ & outgassed $P_\mathrm{CO_2}$  \\
		(kg/kg) & (kg/kg) & (bar) & (bar)  \\
		\hline
		$1.0 \cdot 10^{-4}$ & $1.0 \cdot 10^{-5}$ & 3.3   & 0.6\\
		$1.0 \cdot 10^{-3}$ & $1.0 \cdot 10^{-4}$ & 42.0  & 5.6\\
		$2.0 \cdot 10^{-3}$ & $2.0 \cdot 10^{-4}$ & 85.0  & 11.1\\
		$1.0 \cdot 10^{-2}$ & $1.0 \cdot 10^{-3}$ & 481.6 & 55.5\\   	
		\hline
		\multicolumn{4}{c}{Global magma ocean (1700 km)} \\
		\hline
		mantle [H$_2$O]$_0$ & mantle [CO$_2$]$_0$ & outgassed $P_\mathrm{H_2O}$ & outgassed $P_\mathrm{CO_2}$  \\
		(kg/kg) & (kg/kg) & (bar) & (bar)  \\
		\hline
		$1.0 \cdot 10^{-4}$ & $1.0 \cdot 10^{-5}$ & 7.9    & 1.3\\
		$1.0 \cdot 10^{-3}$ & $1.0 \cdot 10^{-4}$ & 99.2   & 12.8\\
		$2.0 \cdot 10^{-3}$ & $2.0 \cdot 10^{-4}$ & 207.0  & 25.7\\
		$1.0 \cdot 10^{-2}$ & $1.0 \cdot 10^{-3}$ & 1108.7 & 128.0\\      	     	
		\hline      	
	\end{tabular}
\end{table}

\subsection{Adopted parameters and initial conditions}
\label{sec:init}
We use Eqs.~\ref{eq:esc} and \ref{eq:fi} to calculate the evolution of the outgassed hot steam atmospheres \citep{Abe88, Marcq12, Massol16}. For the embryo properties we adopt the mass and radius of current Mars ($M_\mathrm{Mars}=6.42\times10^{23}$~kg, $r_\mathrm{Mars}=3396$~km). The surface temperatures are in the order of $10^3$~K, i.e. magma oceans. Mesopause heights $z_0$ are in the order of several $10^2-10^3$~km for Mars-like bodies with such atmospheres, with mesopause temperatures $T_0$ comparable to the equilibrium temperatures at the orbits \citep{Marcq12}. Such hot atmospheres are typically well-mixed up to large heights \citep{Abe88, Kasting88, Lupu14}. We adopt a mesopause height of $z_0=1000$~km, which is representative for steam atmospheres above a molten magma ocean \citep{Marcq12, Erkaev14}. Thus, $r_0=r_\mathrm{Mars}+z_0=4390$~km. We assume $\beta=1$ for all calculations, as one needs detailed hydrodynamic modeling to evaluate this parameter. In reality both $r_0$ and $\beta$ are time dependent and decrease when the atmospheric mass decreases significantly and/or the composition becomes dominated by the heavier constituents. Thus we tend to overestimate the escape rates, especially at later evolutionary stages. We adopt a heating efficiency of 15\% which is representative for H-dominated atmospheres of hot Jupiters \citep{Shematovich14}, but similar values were also used in other water loss studies \citep{Kasting83, Chassefiere96, Tian15a}. The impact of all these assumptions is discussed in more detail in Section~\ref{sec:disc}.

We assume that atmospheric escape starts at a system age of $t_0=10$~Myr. The gas disk should be dispersed within a few Myr \citep{Haisch01, Montmerle06} and all embryos and existing protoplanets are then exposed to the EUV emission of the host star. Due to the low gravity, any accreted H/He envelope around a Mars-like body is removed very fast on timescales of $<1$~Myr \citep{Erkaev14, Stoekl15, Massol16}. After colliding with other objects, magma oceans may form and degas volatiles during the solidification phase. This occurs on a timescale in the order of 0.1~Myr \citep[e.g.][]{Elkins-Tanton08a, Lebrun13}, depending on orbital distance, planet size, atmospheric mass and composition. Based on the model described in Section~\ref{sec:atm} the outgassing takes 0.5~Myr for the 233~bar and 2~Myr for the 1237~bar ($P_\mathrm{H_{2}O}=1109$~bar, $P_\mathrm{CO_{2}}=128$~bar) atmosphere. The greenhouse effect and thermal blanketing of the outgassed volatiles themselves prolong the magma ocean lifetime. We assume that the initial outgassing occurs close to the age of 10~Myr so that the initial steam atmosphere already exists at this point. Note that our results do not strongly depend on the exact value of $t_0$ because of the initially saturated EUV emission (Section~\ref{sec:euv}). However, new magma ocean phases at later stages are possible after large impacts when the stellar EUV fluxes are already lower compared to the initial stage shortly after disk dispersal.

We study two cases in detail, namely a moderately massive atmosphere with 96~bar ($P_\mathrm{H_{2}O}=85$~bar, $P_\mathrm{CO_{2}}=11$~bar) and a more massive atmosphere with 233~bar ($P_\mathrm{H_{2}O}=207$~bar, $P_\mathrm{CO_{2}}=26$~bar) (cf.~Table~\ref{tab:atm}). Atmospheres of lower masses are removed very quickly. The surface pressures are used to calculate the initial inventories of each species, assuming complete dissociation of H$_2$O into H and O in the upper atmosphere due to the stellar EUV flux. For CO$_2$ we study two limiting cases, one where it remains in molecular form and another where it is completely dissociated into C and O, the latter being likely more realistic for the considered EUV fluxes.

\subsection{Atmospheric loss from Mars-sized embryos}
\begin{figure}
\includegraphics[width=\textwidth]{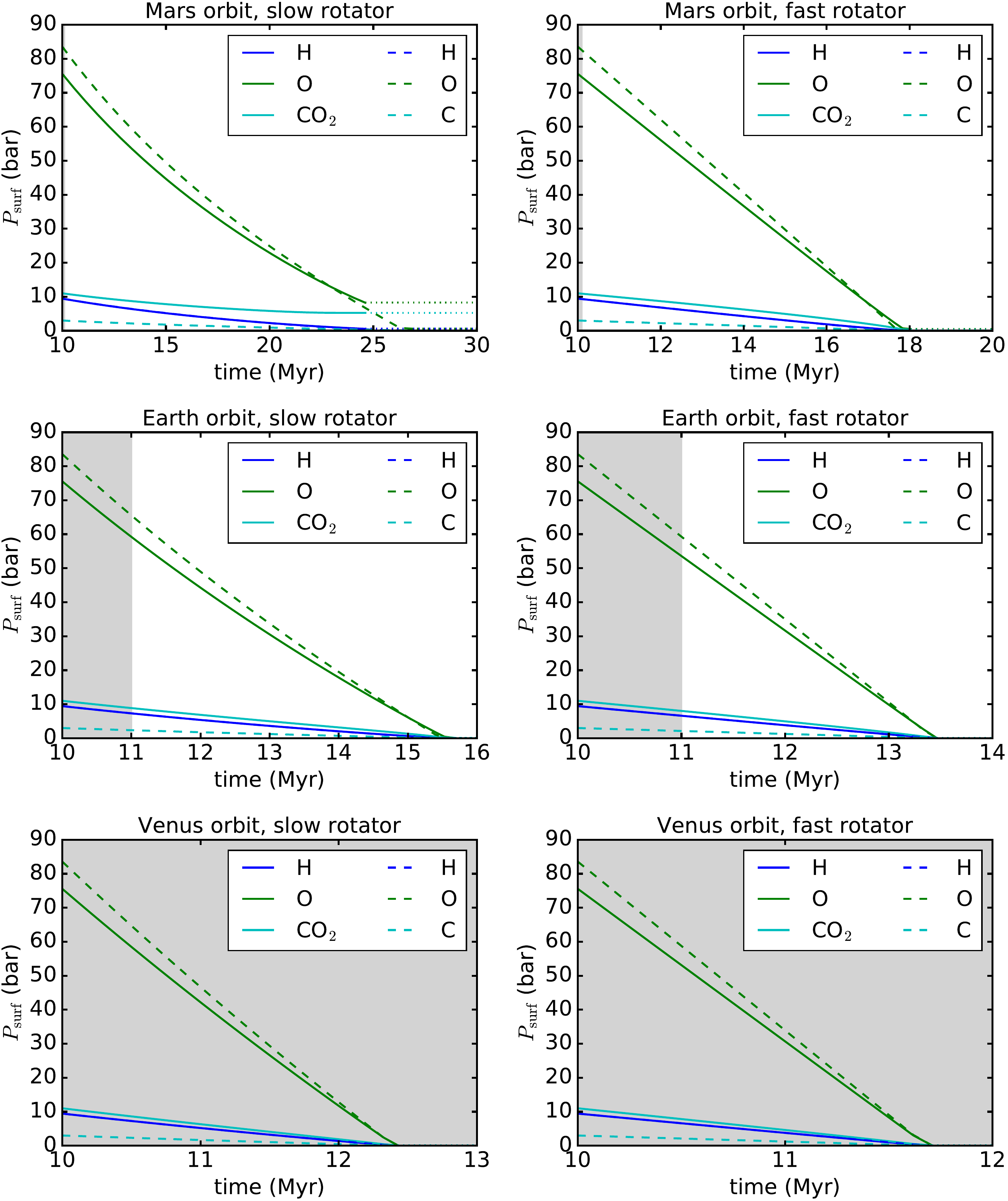}
\caption{Evolution of the partial surface pressures of an escaping steam atmosphere (85~bar H$_2$O, 11~bar CO$_2$) around a Mars-like planetary embryo located at the orbits of Mars, Earth and Venus (top to bottom) around a slow (left) and a fast rotator (right). Solid lines indicate the results if CO$_2$ remains in molecular form and dashed lines assume complete dissociation into O and C. The evolution of H in both cases is almost indistinguishable. Dotted horizontal lines indicate the presence of three major species due to accumulation of a former minor one which cannot be modeled with our present approach. The typical water condensation times at the respective orbits \citep{Lebrun13} are shown as shaded areas (the short 0.1~Myr interval for Mars' orbit is almost invisible).}
\label{fig:mars1}
\end{figure}

\begin{figure}
\includegraphics[width=\textwidth]{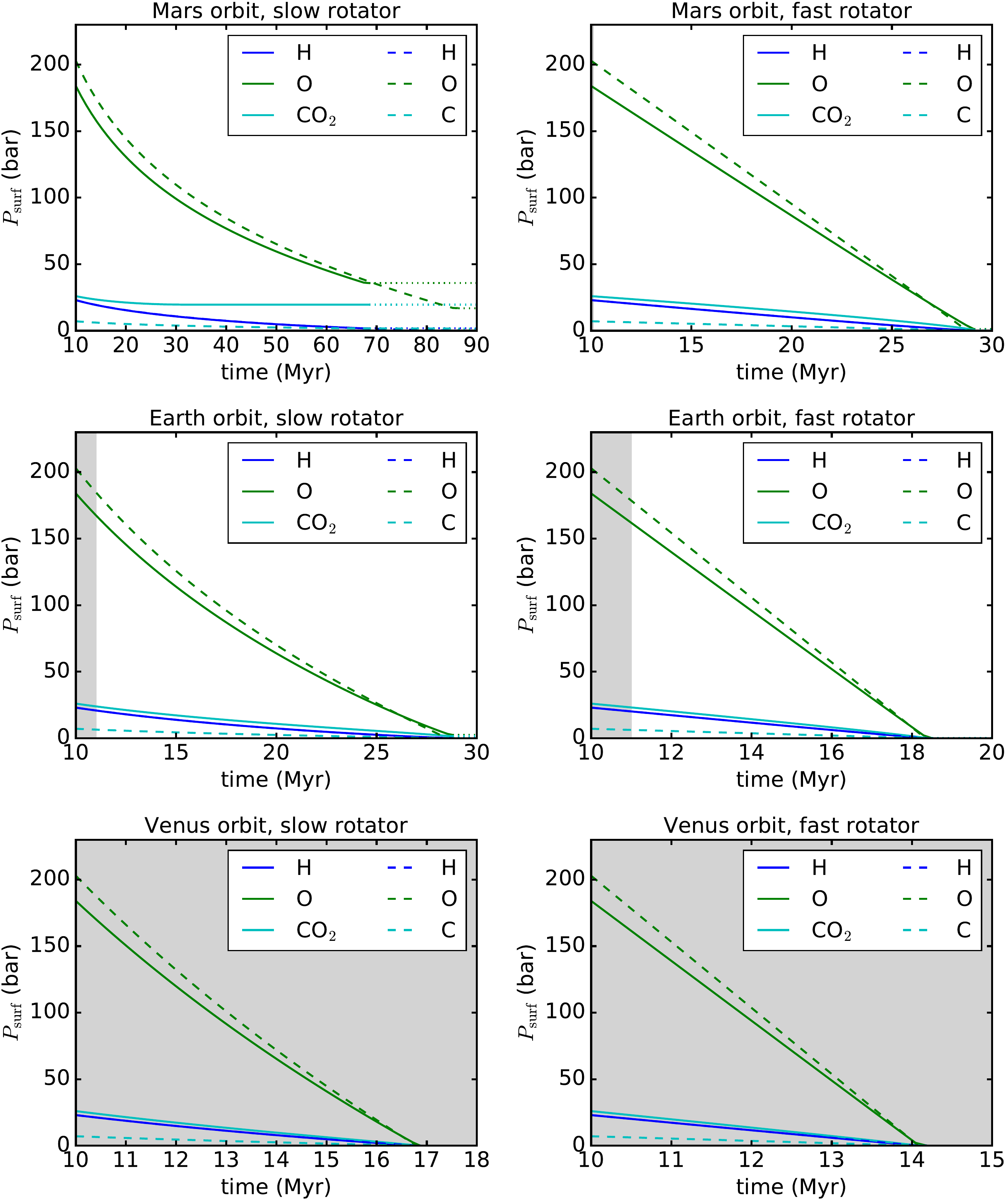}
\caption{Same as Fig.~\ref{fig:mars1}, but for 207~bar H$_2$O and 26~bar CO$_2$.}
\label{fig:mars2}
\end{figure}

Here we present the results of our model of escaping H$_2$O/CO$_2$ steam atmospheres from Mars-like planetary embryos. The resulting evolution of partial surface pressures at different orbits and assuming three stellar rotation tracks are shown in Fig.~\ref{fig:mars1} for the 96~bar case and in Fig.~\ref{fig:mars2} for the 233~bar case.

Generally, the loss of all three species occurs at approximately the same times. Under the assumption of complete dissociation of CO$_2$, C is dragged with the flow efficiently in most cases because of its low mass. However, even CO$_2$ is lost so efficiently that the timescales for atmospheric loss are approximately equal, except for the slow rotator/Mars orbit case. In this scenario, the evolution cannot be followed to the end for CO$_2$ (also for C in the 233~bar case) because the third (initially minor) species accumulates so that the atmosphere consists of three major species (which we define as $N_i/N_\mathrm{tot}\ge0.1$ here). Since there exists no simple analytic approximation to this scenario, such cases would have to be studied using a more advanced numerical model. In case of completely dissociated CO$_2$, the 96~bar atmosphere disappears at 12--27~Myr. For the 233~bar case complete loss occurs at 14--30~Myr, but the slow rotator/Mars orbit case cannot be calculated to the end because at about 85~Myr, C has accumulated. In Fig.~\ref{fig:mars2}, this is represented by subsequent constant evolution of the partial surface pressures indicated by dotted lines. The shaded areas denote the water condensation times as obtained by \citet{Lebrun13}. These are the average times at which the steam atmospheres would condense at the respective orbits according to their model, preventing further escape of H$_2$O. However, this does not take into account that the surfaces are likely to remain molten for longer times due to frequent impacts \citep{Debaille07, Maindl15}, which prolongs the time periods during which water remains in steam form and does not condense. Figures~\ref{fig:mars1} and \ref{fig:mars2} indicate that at Mars' orbit embryos remain rich in water and volatiles.

Under the assumption that all CO$_2$ remains in molecular form, it escapes slightly less efficiently than if dissociated due to its higher mass. Interestingly, the loss times are essentially similar to those for dissociated CO$_2$. This is because the total mass-loss rate $\dot{m}$ (Eq.~\ref{eq:mdot}) remains the same, but the particle escape fluxes $F_{i,j,k}$ of the individual species are different. However, for the slow rotator/Mars orbit case, CO$_2$ escapes so inefficiently that it accumulates so that the atmosphere eventually consists of three major components at 25 (70)~Myr for the 96 (233)~bar atmosphere. As mentioned before, such cases cannot be described with the simple analytic formulae from Section~\ref{sec:escape} \citep[as discussed in][]{Zahnle90} and is complicated by the fact that large amounts of (almost) non-escaping CO$_2$ could substantially reduce the escape of H and O which must then diffuse through a static background atmosphere. Moreover, the presence of non-negligible amounts of CO$_2$ could lead to efficient cooling by infrared radiation \citep[e.g.][]{Kulikov07}.

The maximum possible outgassed atmosphere of 1237~bar (Table~\ref{tab:atm}) is removed at 32--55~Myr between Venus' and Earth's orbit in the fast rotator case and at 35~Myr in the moderate rotator case at Venus' orbit. All other combinations of orbits and rotation tracks would not be able to remove such a massive atmosphere within 100~Myr. On the other hand, the low-mass atmospheres from Table~\ref{tab:atm} are removed very quickly in all orbits for all rotation tracks, typically within a few Myr.

The efficient loss of heavier atoms like O and C and even molecules like CO$_2$ also has consequences for noble gases and their isotopes which may also be dragged with the outflow of the main atmospheric constituents. Their escape efficiency, or if they escape at all, depends on their masses. Lower mass elements escape more efficiently than elements with higher masses (cf.~Eq.~\ref{eq:xk}). If an element with a certain mass is dragged with the flow this implies that elements with lower masses also escape. For instance, rapid loss of O$_2$ would imply that also Ne and its isotopes are removed quickly. If CO$_2$ escapes, also Ar and its isotopes are removed. Therefore, since larger terrestrial planets likely form from the collisions of smaller bodies with masses similar to that of Mars, and since such smaller planets can experience fast atmospheric mass-loss, volatiles in the atmospheres of planetary embryos might already be fractionated before the embryos are incorporated into planets. However, this is only relevant if embryos still possess remnant atmospheres when they collide. Moreover, since Mars-like embryos lose their volatiles within a few Myr at the orbit of Venus, it is likely that the building blocks of Venus were volatile-poor, especially if it accreted late. However, if some embryos were scattered inwards from regions beyond Mars' orbit, these could have provided volatile-rich material. This may have affected the formation and early evolution of protoplanets.

\section{Atmospheric loss from Proto-Venus}\label{sec:venus}
\begin{figure}
	\includegraphics[width=\columnwidth]{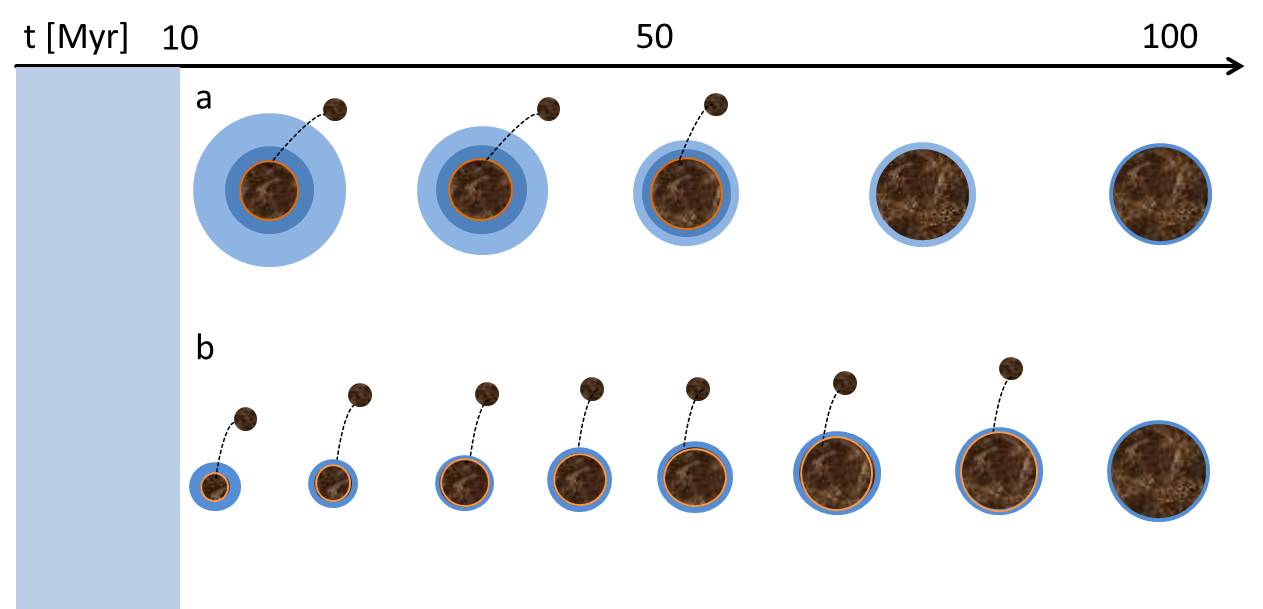}
	\caption{Sketch of possible evolution scenarios of Venus. (a) A proto-Venus formed within the nebula and accreted a H-dominated protoatmosphere. Large impacts contributed to the final planet mass. (b) Only embryos were present in the protoplanetary nebula, the final planet forms through collisions between embryos.}
	\label{fig:sketch}
\end{figure}

To study the effect of the previously discussed issues on the formation of protoplanets we investigate the early evolution of Venus as an example. Two possible scenarios of Venus' past evolution are sketched in Fig.~\ref{fig:sketch}. The first possibility is an early accretion of a proto-Venus within the nebula, which results in accumulation of a hydrogen-dominated protoatmosphere. That proto-Venus and proto-Earth grew to masses of about $0.5{-}0.75M_{\oplus}$ within 10~Myr so that they could capture thin H$_2$-envelopes is also in agreement with the model results for the ``Grand Tack'' hypothesis \citep{Walsh11, OBrien14}. These recent findings on dynamics, terrestrial planet formation, as well as observations of low-mass exoplanets with H-dominated envelopes, support earlier hypotheses that the early Earth may have been surrounded by a thin H$_2$-dominated protoatmosphere. Isotope fractionation may then be obtained from escape of this protoatmosphere with possible additionally outgassed volatiles. The second possibility is late accretion where the planet formed after the nebula dissipated. In this case fractionation can only be obtained by escape of a steam atmosphere. However, in this scenario the question arises how volatile-rich such a planet may be since both nebula-accreted and outgassed atmospheres are lost from planetary embryos efficiently, especially at the orbit of Venus \citep[cf. Section~\ref{sec:embryos};][]{Albarede07, Albarede09}.

Possible scenarios are constrained by observations of noble gas isotope ratios \citep[e.g.][]{Gillmann09}. The fractionation of noble gases in an escaping atmosphere is described by Eq.~\ref{eq:xk} and can be easily included in the model. Since these elements are very minor compared to the other atmospheric constituents, one can ignore their impact on the escape rates of the more prominent species (Eq.~\ref{eq:mdot}) and simply calculate their escape fluxes using Eq.~\ref{eq:esc}. We assume that initially the abundances of the noble gases in the atmosphere are equal to the solar values. To estimate the initial amounts of Ne, Ar and their isotopes, we use the solar abundances given in \citet{Halliday13} and the solar isotope ratios from \citet{Pepin91}. We determine all abundances relative to H \citep{Halliday13} and multiply these ratios with the initial atmospheric H inventories. Note that the evolution of an isotope ratio of a single element is independent of its initial abundance. Binary diffusion parameters for Ne and Ar are taken from \citet{Zahnle86} and are assumed to be equal for an element and its isotopes.

\subsection{Early accretion}\label{sec:early}
\subsubsection{Adopted parameters and initial conditions}
In the early accretion scenario it is assumed that a proto-Venus has grown to a mass of $0.7~M_\mathrm{Venus}$ within the protoplanetary nebula. We assume the protoplanet's radius is $r_p=0.89~R_\mathrm{Venus}$, yielding the same density as present Venus. The missing mass is expected to be delivered within $\le50$~Myr by impacts of planetesimals and planetary embryos. As before, we adopt a starting time of $t_0=10$~Myr and assume that initial atmospheric escape due to ``boil-off'' has already occurred before this time. Moreover, since escape rates in the boil-off regime are very high compared to the EUV-driven regime we can assume that it does not fractionate the considered isotopes efficiently before the slower EUV-driven escape sets in. Thus we only consider remnant envelope mass fractions $<$3\% at $t_0$ \citep{Stoekl16}. Moreover, we assume that also an outgassed Venus-type steam atmosphere of 559~bar ($P_\mathrm{H_{2}O}=458$~bar, $P_\mathrm{CO_{2}}=101$~bar) \citep{Lichtenegger16} is already present. We note, however, that an assumed steam atmosphere only marginally affects the results, especially for high H contents.

The choice of $r_0$ for such H-rich atmospheres has a strong influence on the results. According to models \citep[e.g.][]{Lopez14} a young protoplanet with $0.7M_\mathrm{Venus}$ may have an optical transit radius up to $\sim4.7R_\oplus$ ($\sim5R_\mathrm{Venus}$) for a 1\% H-dominated atmosphere at an age of 10~Myr. This would place the protoplanet in the boil-off regime. Since we consider the slower EUV-driven evolution only, we estimate an upper limit to the radius ($r_\mathrm{max}$) by requiring that the escape parameter at $r_0$, $\lambda_0=GM\mu m_H/(kT_0r_0)$, must be $\ge20$ \citep{Fossati17}. We adopt $\mu=2.35$ as for solar composition gas (neglecting the steam contribution because the considered nebula atmospheres are typically much more massive). As a minimum radius ($r_\mathrm{min}$) we assume that of a pure steam atmosphere $r_p+300$~km \citep{Lichtenegger16}. Note that we keep $r_0$ fixed in time as before, but study the influence of different choices of $r_0$.

We explore two different scenarios, one with a proto-Venus ($0.7M_\mathrm{Venus}$, $0.89R_\mathrm{Venus}$) and three Mars-like impactors, and one with a fast-forming complete Venus ($M_\mathrm{Venus}=4.87\times10^{24}$~kg, $R_\mathrm{Venus}=6052$~km) emerging out of the nebula to illustrate the influence of the assumptions about the initial condition and impact history. We obtain $r_\mathrm{max}=2r_p=1.78R_\mathrm{Venus}$ and $2.55R_\mathrm{Venus}$ for masses of $0.7M_\mathrm{Venus}$ and $1M_\mathrm{Venus}$, respectively. The adopted minimum radii correspond to $1.06r_p=0.94R_\mathrm{Venus}$ and $1.05R_\mathrm{Venus}$, respectively. Here we assume that the impactors have $0.75M_\mathrm{Mars}$ and $0.9R_\mathrm{Mars}$ so that the final planet has $1M_\mathrm{Venus}$, for better comparison between the two scenarios. The impactors only add mass to the protoplanet's core. We add three impactors at 20, 30 and 40~Myr, to finally obtain a ${\sim}1M_\mathrm{Venus}$ planet.

\subsubsection{Atmospheric evolution}
To study the EUV-driven atmospheric evolution and associated fractionation of noble gases, we calculate a grid of evolutionary scenarios with different initial H inventories, a steam atmosphere (assumed to be well mixed with H) and the three stellar rotational evolution tracks. Since our results do not depend significantly on the assumption of molecular or dissociated CO$_2$ because the initial partial pressures of H are much larger than that of the other steam constituents, we keep it in molecular form.

The evolution of the noble gas isotope ratios depends sensitively on the escape rates (and their evolution), as well as the initial atmospheric H content. For all three stellar rotation tracks we aim to find the initial amount of H with which we can reproduce the present-day isotope ratios at some point during the evolution. Moreover, we require that the isotope ratios remain constant within the uncertainties of the measurements to ensure that they are not overfractionated later. At t$_0$, we adopt a $\sim$560~bar steam atmosphere and add additional H (assumed to be a remnant of the accreted protoatmosphere from nebula), leaving the exact amount as a free parameter.  Generally, the isotope ratios evolve from values higher than on present Venus (i.e. solar ones) towards the desired parameter region, but may also drop below these values. If too much H is present at $t_0$, the ratios always remain above the present-day values during the evolution; if too little H is assumed, they may drop below present-day values. Only small ranges of H abundances at $t_0$ lead to the desired values of the isotope ratios for every considered case.

Table~\ref{tab:iso} gives the range of H mass fractions from the nebula which lead to isotope ratios in the range of present-day Venusian values. We find valid solutions for almost all scenarios, except for the fast rotator/$r_\mathrm{max}$ case for which unrealistically large amounts of H ($\gg$3\%) would have to be added. In the slow rotator/$r_\mathrm{min}$ cases, the ratios remain slightly above present-day levels even if H is reduced to zero and pure steam atmospheres are assumed. Table~\ref{tab:iso} also gives the system ages $t_\mathrm{H}$ at which all hydrogen would be lost (i.e. its partial pressure drops below a few bar) and the remaining amount of CO$_2$ at these times. The latter is important because Venus has about 90~bar of CO$_2$ in its atmosphere; therefore scenarios reducing CO$_2$ by a large amount are less realistic. Generally, scenarios with more efficient escape (e.g. due to faster stellar rotation or larger assumed $r_0$ values) require more additional H and therefore have longer evolution times $t_\mathrm{H}$. This is because if escape rates are high, isotopes of a single element with their comparable masses are dragged with the flow at more similar rates which decreases fractionation. This means that longer evolution times are needed until the present isotope ratios are reached, which is only possible if the initial atmospheres are massive enough.

\begin{figure}
	\includegraphics[height=9cm]{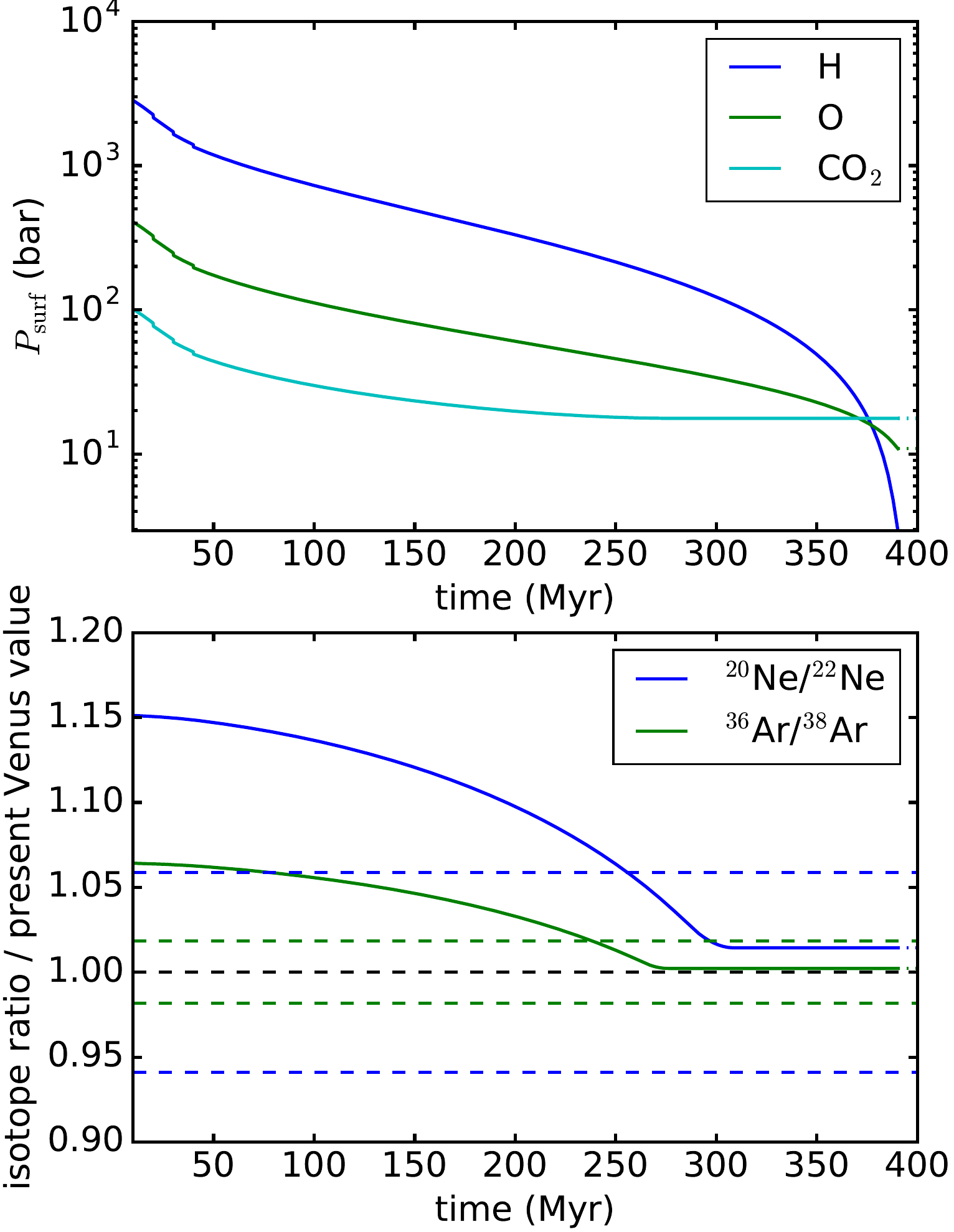}
	\includegraphics[height=9cm]{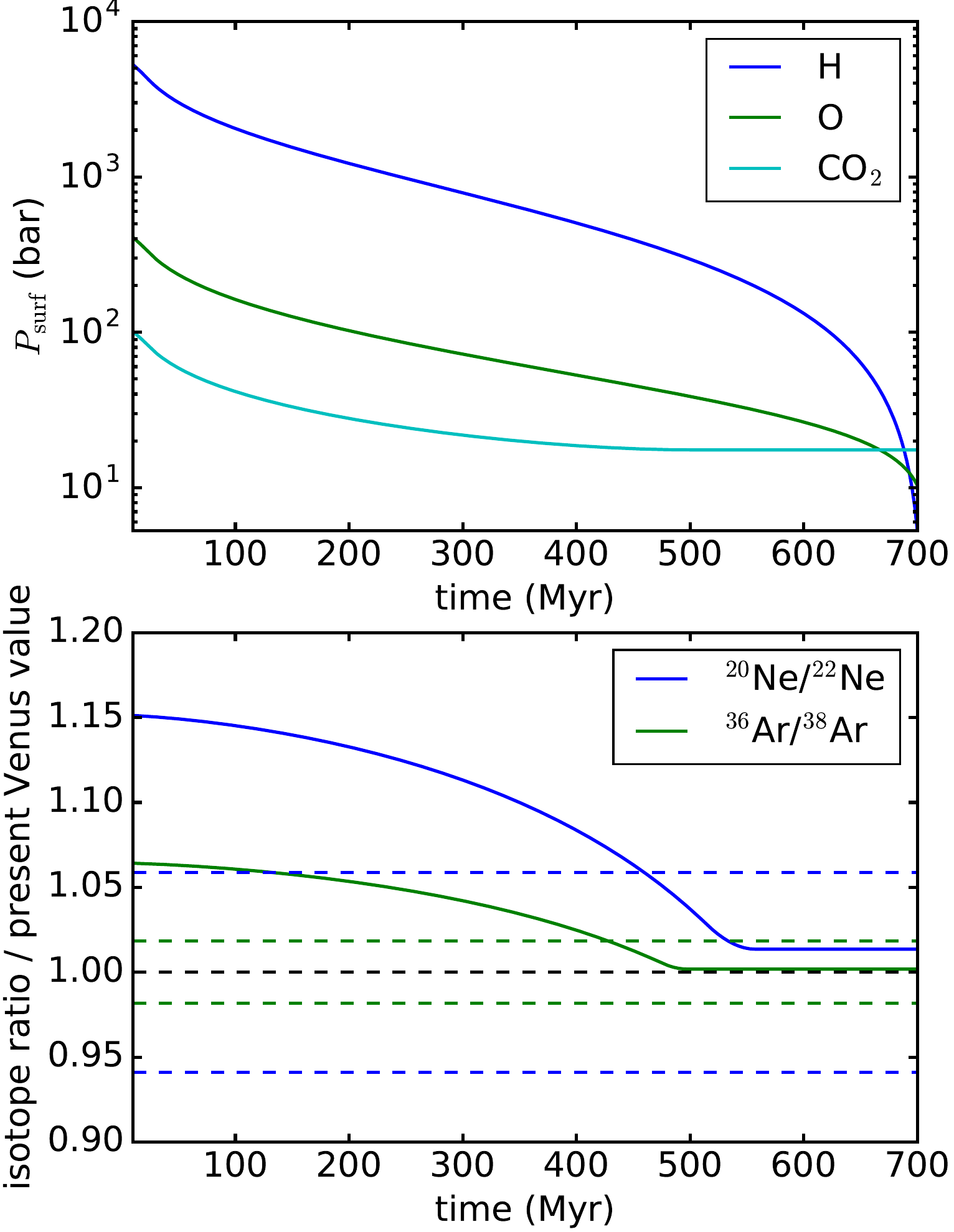}
	\caption{Evolution of a mixed nebula and 559~bar steam atmosphere of a proto-Venus with impacts (left) and a complete Venus (right). The left panels show the case of a moderate rotator, $r_0=0.8r_\mathrm{max}$, and 0.38\% of accreted H. The right panels show the case of a moderate rotator, $r_0=0.7r_\mathrm{max}$, and 0.556\% of accreted H. Upper panels show the atmospheric evolution, lower panels the evolution of the noble gas isotope ratios. Dashed lines give the ranges of the presently measured Venusian values.}
	\label{fig:venusisoneb}
\end{figure}

\begin{figure}
	\includegraphics[height=9cm]{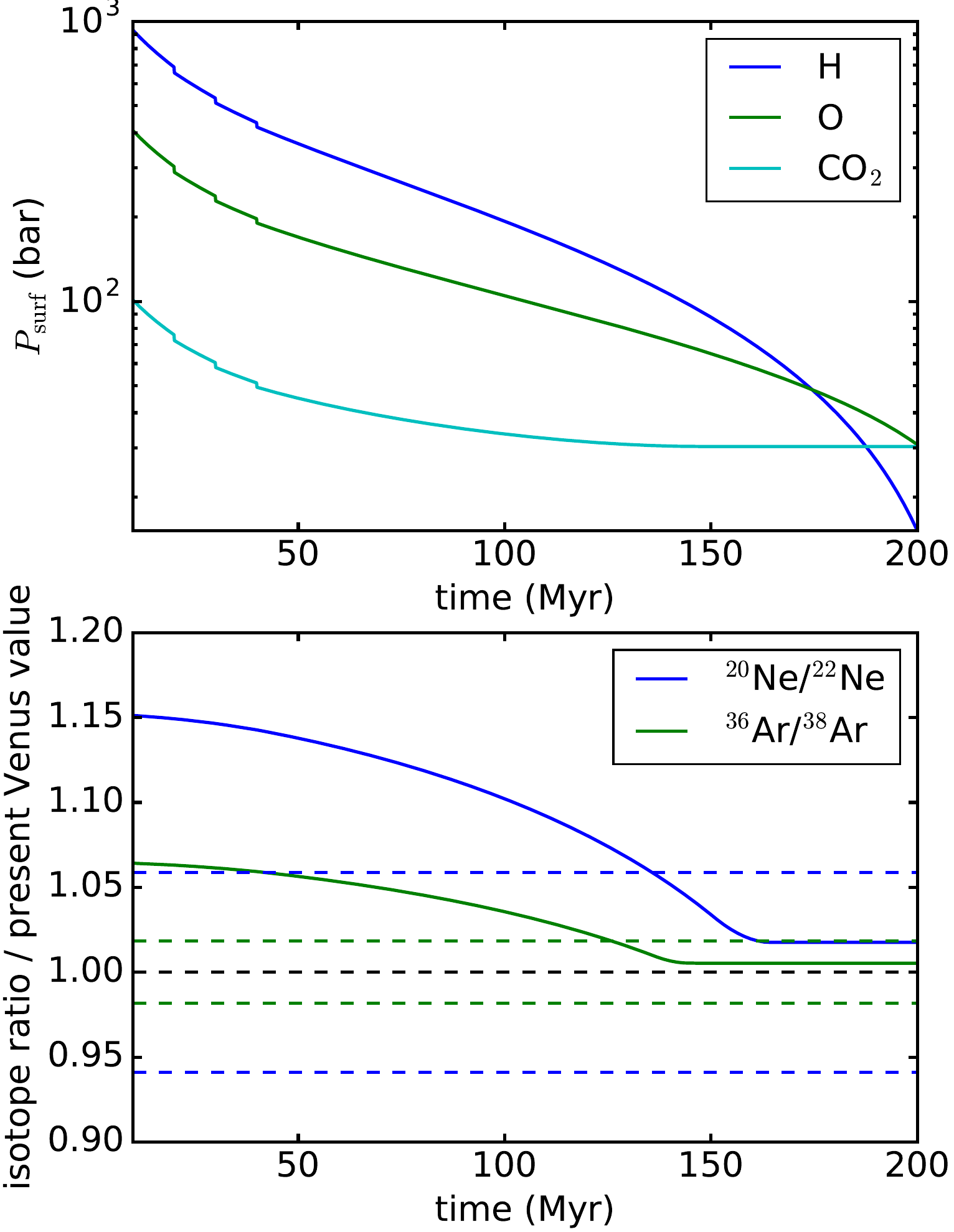}
	\includegraphics[height=9cm]{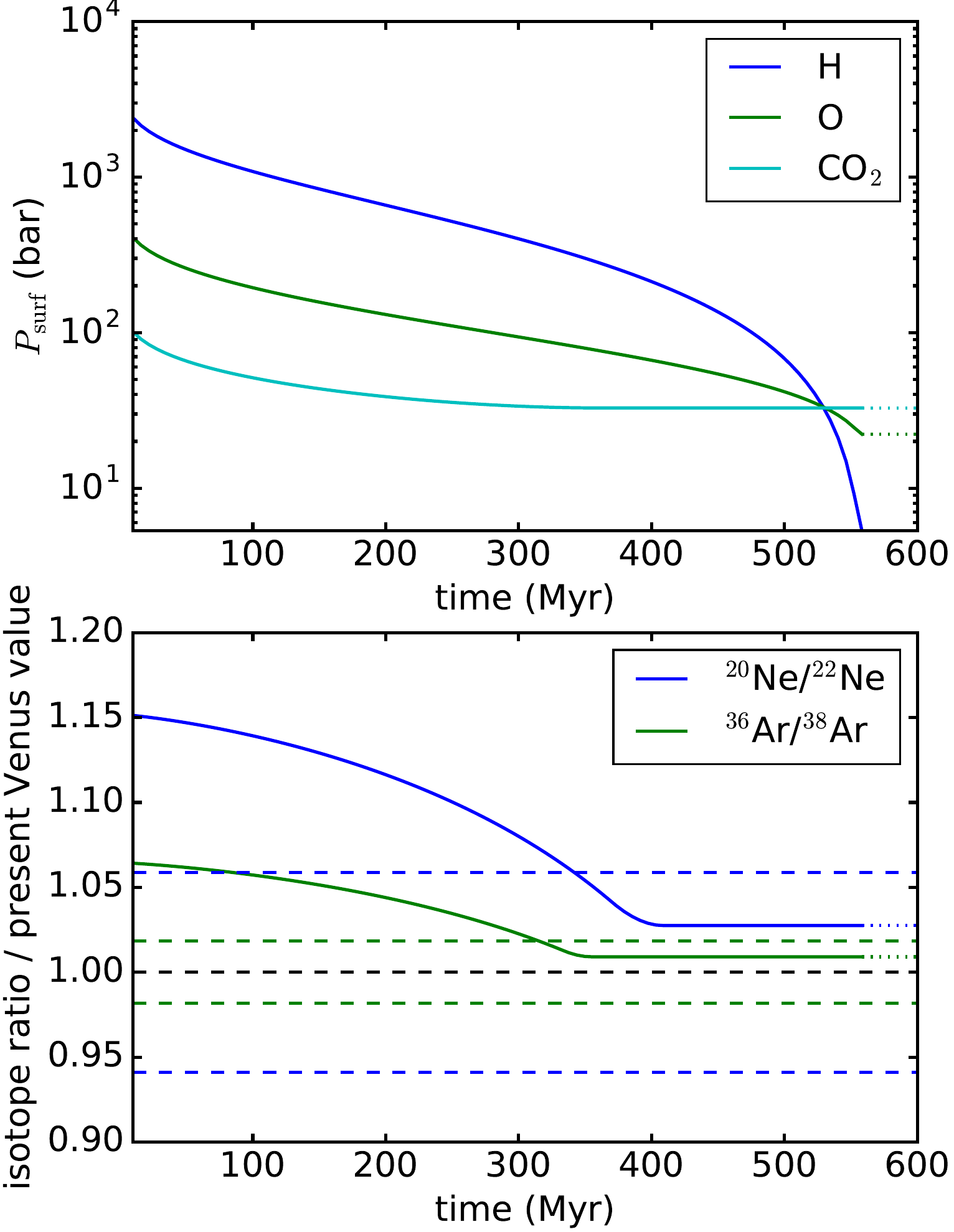}
	\caption{Same as Fig.~\ref{fig:venusisoneb}, but for a slow rotator and assuming $r_0=0.8r_\mathrm{max}$ with 0.12\% of accreted H (for the proto-Venus, left panels), and $r_0=0.7r_\mathrm{max}$ with 0.25\% of accreted H (for the complete Venus, right panels).}
	\label{fig:venusisonebslow}
\end{figure}

Examples yielding present-day Venusian isotope ratios are shown in Figs.~\ref{fig:venusisoneb} and \ref{fig:venusisonebslow}. In Fig.~\ref{fig:venusisoneb} the proto-Venus example shows a moderate rotator/$0.8r_\mathrm{max}$ case with 0.38\% H and the Venus example a moderate rotator/$0.7r_\mathrm{max}$ case with 0.556\% of H. Note that such H envelopes can be removed using more sophisticated hydrodynamic escape models within similar timescales \citep{Johnstone15a}. For a slow rotator, Fig.~\ref{fig:venusisonebslow} shows cases of $0.8r_\mathrm{max}$ with 0.12\% H (proto-Venus) and $0.7r_\mathrm{max}$ with initially 0.25\% of H (complete Venus). The isotope ratios are normalized to the present values of $^{20}\mathrm{Ne}/^{22}\mathrm{Ne}=11.9\pm0.7$ and $^{36}\mathrm{Ar}/^{38}\mathrm{Ar}=5.45\pm0.1$ \citep{Gillmann09} and the measurement uncertainties are indicated by dashed lines of the same color as the corresponding modeled ratios. Generally we find that for faster rotating stars (i.e. higher EUV flux levels) more H from the nebula is required to produce the desired isotope ratios, similar to the scenario with pure steam atmospheres. The valid solutions we find all lie in a very restricted parameter space which does not allow us to select a most favorable case. However, since we find at least some possible range in almost every evolution history, this indicates that it is likely that thermal escape of a remnant accreted nebula atmosphere is a likely cause of the presently observed isotope ratios on Venus. This also indicates that it is possible that a sufficiently large proto-Venus was already formed when the nebula was still present. A similar scenario has been suggested for Earth where Xe isotopes may be explained by an accreted protoatmosphere which was mixed into the upper regions of the mantle \citep{Becker03}.

Note that we are not able to reproduce Venus' present $^{20}\mathrm{Ne}/^{36}\mathrm{Ar}$ abundance ratio if initially starting from solar abundances, which always remains close to solar. It has been recognized in a previous application of such a model to Mars that it is not possible to obtain this abundance ratio simultaneous with the isotope ratios starting from solar values \citep{Zahnle90}. If aiming to reproduce this ratio, the isotope ratios would be overfractionated. This is discussed in more detail in Section~\ref{sec:disc}.

The impact history has some influence on the required nebula H and timescales, but not on the general conclusion that hydrodynamic escape can lead to the presently observed noble gas isotope ratios. We also compare a Venus case without a steam atmosphere and only H from the nebula to evaluate the influence of the presence of the steam atmosphere at $t_0$. We also find parameter ranges which may reproduce the presently observed isotope ratios. The time at which H is lost is comparable to the case with an added steam atmosphere; however, small differences occur for smaller $r_0$ and a slowly rotating host star. The required H at $t_0$ is higher than for the steam+nebula case, which is clear since the amount of atmosphere needed to obtain the final isotope ratios must be comparable for fixed stellar evolution tracks. This indicates that the results are not strongly dependent on the assumed steam atmosphere, since no steam simply raises the required nebula H at $t_0$. Another important aspect is that both Ne and Ar isotope ratios reach the present-day ranges at roughly similar times for all cases, indicating that hydrodynamic escape is a likely cause of the observed fractionation of these noble gas isotopes.

It is important to note that the numbers given in Table~\ref{tab:iso} are only valid for $T_0=300$~K, which is the adopted equilibrium temperature at the orbit of Venus. There is some dependence on the choice of the temperature \citep[cf. also][]{Gillmann09}. Increasing $T_0$ leads to more efficient dragging because this increases $x_{j,k}$ (cf.~Eqs.~\ref{eq:xj} and \ref{eq:xk}), and therefore also $F_{j,k}$. The H amounts required to reach valid isotope ratio ranges therefore increases, as do the corresponding loss times. 

If a steam atmosphere was already present at $t_0$ and mixed with residual H, our results indicate that a slow rotator and/or smaller $r_0$ produce favorable results because otherwise too much CO$_2$ would escape. However, if the outgassing occurred later when the nebula atmosphere was sufficiently tenuous the isotope ratios could possibly also be reproduce via mixing the fractionated residual H with a newly outgassed steam atmosphere and subsequent escape thereof, provided that the host star remains active for a sufficiently long time.

From the results in Table~\ref{tab:iso}, one can see that the remaining CO$_2$ contents are rather low which restricts possible scenarios further. There are several ways in which substantial loss of CO$_2$ could be mitigated. The modeling results point towards rather low H contents at $t_0$, small $r_0$, as well as a low-activity host star. Furthermore, since the results are shown for CO$_2$ remaining in molecular form which makes escape less efficient, a high fraction of dissociated CO$_2$ is also unlikely. Since being exposed to the strong EUV fluxes of a rapidly rotating star CO$_2$ is very likely to be dissociated in the escape region; this also suggests that the young Sun had a low level of activity, corresponding to it being a slow rotator. Another possibility is that the first outgassed steam atmosphere was more massive and had CO$_2$ contents $\gg$100~bar. We made some tests with a more massive 770~bar ($P_\mathrm{H_{2}O}=630$~bar, $P_\mathrm{CO_{2}}=140$~bar) steam atmosphere. In the  proto-Venus case with a moderate rotator and $r_0=r_\mathrm{min}$, about 60~bar CO$_2$ are left when the isotope ratios reach present values and for $0.8r_\mathrm{max}$, 30~bar remain. For a slow rotator and $0.8r_\mathrm{max}$, 50~bar are left. These results indicate that simply raising the amount of CO$_2$ at $t_0$ by a certain amount does not lead to residual CO$_2$ pressures higher by the same amount, but by a lower value. Note that H contents different from those of Table~\ref{tab:iso} had to be added to reach present isotope ratios.

A further solution could be that more CO$_2$ was delivered later by impacts. In this case, the impactors would have had to have kept their volatiles until the time that they impacted with the planet; one way they could have done this is by being small enough to not have a magma ocean, so that the volatiles would only be released from the body during the impact. According to \citet{Elkins-Tanton08} and \citet{Elkins-Tanton11} more than 85\% of the H$_2$O and more than 90\% of the CO$_2$ contents of a magma ocean are degassed into the atmosphere on an Earth-mass planet. However, if either the magma ocean did not involve the entire mantle and/or was not global, a large volatile fraction could have been retained in the interior, even if almost the whole volatile content of the magma ocean is degassed upon solidification. This effect could also inhibit the efficient volatile loss from smaller embryos discussed in Section~\ref{sec:embryos}.  This indicates that since the initial magma ocean, if it was global and deep, releases most of its volatiles the crucial point is likely the time when this occurs. It is also possible that the initial outgassing only occurred after most H was already gone since the large surface pressures of the nebula-accreted atmosphere could have prevented degassing and kept the surface temperatures at sufficiently high levels that the magma ocean solidification was delayed. The delayed outgassing would mean that by the time the steam atmosphere formed, the Sun's EUV emission could have already decayed to lower values and there would be less H present to drag CO$_2$ with it. A late veneer of chondritic material, similar as suggested for Earth, could have provided additional volatiles at 70--140~Myr \citep{Albarede09}. However, if in this scenario also 1--2 Earth oceans of water were delivered, it is possible that large amounts of oxygen could have remained on the planet. Some amount of CO$_2$ could also have been outgassed over geologically long times via volcanism.

Another issue in this context is the timescale of atmospheric loss. While it is essentially unconstrained at Venus, we know that Earth already had liquid water on its surface at an age of 200--300~Myr \citep{Mojzsis01}, indicating atmospheric pressures much lower than those of massive protoatmospheres. Since both planets may have had a broadly similar evolution, this would indicate that timescales of $<300$~Myr for the loss of any type of massive protoatmosphere are favored. For instance, the scenario shown in the left panel of Fig.~\ref{fig:venusisonebslow} would be consistent with such an evolution. The missing CO$_2$ would have to be degassed later by resurfacing events, super-volcanism, or delivered by impacts. One can see from Table~\ref{tab:iso} that low H contents at 200--300~Myr occur only in 2--3 scenarios. In scenarios with longer atmospheric loss timescales and moderate CO$_2$ loss, the partial surface pressures of H exceed a few 100~bar at this age. Hydrogen may also have an important contribution to the greenhouse effect \citep{Pierrehumbert11, Wordsworth12} which may keep surface temperatures high for extended timescales.

\begin{table}
	\caption{Required nebula H inventories to reach present Venusian Ne and Ar isotope fractionation patterns in presence of a 559~bar steam atmosphere, system age $t_\mathrm{H}$ at which H is lost, and residual $P_\mathrm{CO_2}$ at this age.}
	\label{tab:iso}
	\centering
	\begin{tabular}{ccccc}
		\hline
		\multicolumn{5}{c}{Proto-Venus (steam+nebula+impacts)} \\
		\hline
		Rotator & $r_0$ & initial H & $t_\mathrm{H}$ & $P_\mathrm{CO_2}(t_\mathrm{H})$  \\
		  &   & (\%) & (Gyr) & (bar)  \\
		\hline
		Fast & $r_\mathrm{max}$    & --  & -- & -- \\
		Fast & $0.8r_\mathrm{max}$ & 2.476--2.57 & 1.15--1.9 & 6--10 \\
		Fast & $r_\mathrm{min}$    & 0.874--0.94 & 0.73--0.9  & 11--17 \\
		Moderate & $r_\mathrm{max}$    & 0.873--0.945 & 0.6--1      & 11--17 \\
		Moderate & $0.8r_\mathrm{max}$ & 0.363--0.407 & 0.32--0.5   & 14--22 \\
		Moderate & $r_\mathrm{min}$    & 0.055--0.075 & 0.125--0.19 & 34--40 \\
		Slow & $r_\mathrm{max}$    & 0.375--0.44 & 0.43--0.7  & 19--29 \\
		Slow & $0.8r_\mathrm{max}$ & 0.098--0.14 & 0.14--0.28 & 23--35 \\
		Slow & $r_\mathrm{min}$    & -- & -- & -- \\
		\hline      	
		\multicolumn{5}{c}{Venus (steam+nebula)} \\
		\hline
		Rotator & $r_0$ & initial H & $t_\mathrm{H}$ & $P_\mathrm{CO_2}(t_\mathrm{H})$  \\
		  &   & (\%) & (Gyr) & (bar)  \\
		\hline
		Fast & $r_\mathrm{max}$    & -- & -- & -- \\
		Fast & $0.7r_\mathrm{max}$ & 3.426--3.515 & 1.5--2.5 & 4--7 \\
		Fast & $r_\mathrm{min}$    & 0.597--0.646 & 0.7--1.2 & 12--20 \\
		Moderate & $r_\mathrm{max}$    & 1.926--2.04   & 1.5--2.5  & 10--16 \\
		Moderate & $0.7r_\mathrm{max}$ & 0.54--0.593   & 0.6--1    & 15--23 \\
		Moderate & $r_\mathrm{min}$    & 0.0332--0.049 & 0.12--0.2 & 45--50 \\
		Slow & $r_\mathrm{max}$    & 1.06--1.2    & 1.45--2.5 & 20--30 \\
		Slow & $0.7r_\mathrm{max}$ & 0.222--0.276 & 0.4--0.75 & 25--38 \\
		Slow & $r_\mathrm{min}$    & -- & -- & -- \\     	     	
		\hline      	
		\multicolumn{5}{c}{Venus (nebula only)} \\
		\hline
		Rotator & $r_0$ & initial H & $t_\mathrm{H}$ & $P_\mathrm{CO_2}(t_\mathrm{H})$  \\
		&   & (\%) & (Gyr) & (bar)  \\
		\hline
		Fast & $r_\mathrm{max}$    & -- & -- & -- \\
		Fast & $0.7r_\mathrm{max}$ & 3.49--3.59 & 1.5--2.5 & -- \\
		Fast & $r_\mathrm{min}$    & 0.66--0.7  & 0.74--1.2 & -- \\
		Moderate & $r_\mathrm{max}$    & 1.98--2.1    & 1.5--2.5   & -- \\
		Moderate & $0.7r_\mathrm{max}$ & 0.6--0.65    & 0.6--1     & -- \\
		Moderate & $r_\mathrm{min}$    & 0.092--0.105 & 0.17--0.29 & -- \\
		Slow & $r_\mathrm{max}$    & 1.125--1.25  & 1.5--2.5   & -- \\
		Slow & $0.7r_\mathrm{max}$ & 0.293--0.33  & 0.4--0.75  & -- \\
		Slow & $r_\mathrm{min}$    & 0.0332--0.04 & 0.09--0.14 & -- \\     	     	
		\hline      	
	\end{tabular}
	
\end{table}

\subsection{Late accretion}\label{sec:late}
If a body of $0.7M_\mathrm{Venus}$ was already assembled during the circumstellar gas disk phase, it would have accreted some residual gas and started its evolution most likely with a dense H-dominated protoatmosphere, as discussed in Section~\ref{sec:early}. The question remains if cooling, solidification and subsequent volatile degassing of a magma ocean would take place if the atmospheric pressure of an accreted H atmosphere is non-negligible. Since such a proto-Venus would require longer timescales to lose a hydrogen envelope due to its higher gravity than e.g. Mars-like bodies \citep{Erkaev14}, outgassing may occur at a later time. However, there exists the alternative possibility that Venus formed late (at 50--100~Myr) from the assembly of several planetary embryos. In the late accretion scenario, Venus is expected to be formed within 70~Myr from collisions between large planetary embryos and planetesimals where the building blocks all remained smaller than or comparable to Mars within the nebula. We estimate this starting point by assuming that Venus should have been fully formed at a time comparable to the Moon-forming impact at Earth, which is dated between 50 and 150~Myr \citep{Halliday00, Connelly16}, with recent studies pointing towards $\sim$60~Myr \citep{Barboni17}.

We aim to test if the present isotope ratios could also be reproduced in such a scenario. We start our calculations at 70~Myr and assume that at this point, the formation of Venus ($1M_\mathrm{Venus}$, $1R_\mathrm{Venus}$) has finished and that its initial magma ocean has solidified and degassed its volatile contents. However, we do not put constraints on the possible amounts of volatiles here because the planetary embryos from which the planet formed may have already (partly) lost their volatile inventories, depending on when (or if) they were outgassed and how much of the total content was involved. Moreover, a late veneer may have delivered additional volatiles. Thus we only aim to find possible atmospheric amounts and compositions which would allow the reproduction of current isotope ratios and discuss the possible existence of the resulting atmospheres in succeeding cases.

For a starting time of $t_0=70$~Myr, we find no valid solutions for the slow rotator track. The EUV fluxes in this scenario are already so low that both Ne and Ar are not dragged with the flow. In case of a moderate rotator, the present $^{20}\mathrm{Ne}/^{22}\mathrm{Ne}$ can be reproduced in some cases, but Ar is only fractionated up to 85~Myr, which is too short so that the present $^{36}\mathrm{Ar}/^{38}\mathrm{Ar}$ isotope ratio is never reached. We can only find valid solutions for the fast rotator case. Assuming that all CO$_2$ is present in molecular form, the final isotope ratios lie within present ranges for 5000--6000~bar H$_2$O and 110~bar CO$_2$. The CO$_2$ content is in this case not so important; it was chosen mainly such that the final CO$_2$ pressure remains about 90~bar. Massive water envelopes are necessary in order to provide H for a sufficiently long time to reach present isotope ratios. In case of completely dissociated CO$_2$, which is a likely scenario for a fast rotator, the range of possible atmospheric pressures is 4700--5800~bar H$_2$O. The corresponding CO$_2$ contents have to be raised to 350 and 270~bar, respectively, again mainly chosen to leave about 90--100~bar of CO$_2$ on the planet. The choice of CO$_2$ is more sensitive in this case because atomic C is dragged with the flow easier.

\begin{figure}
	\includegraphics[height=9cm]{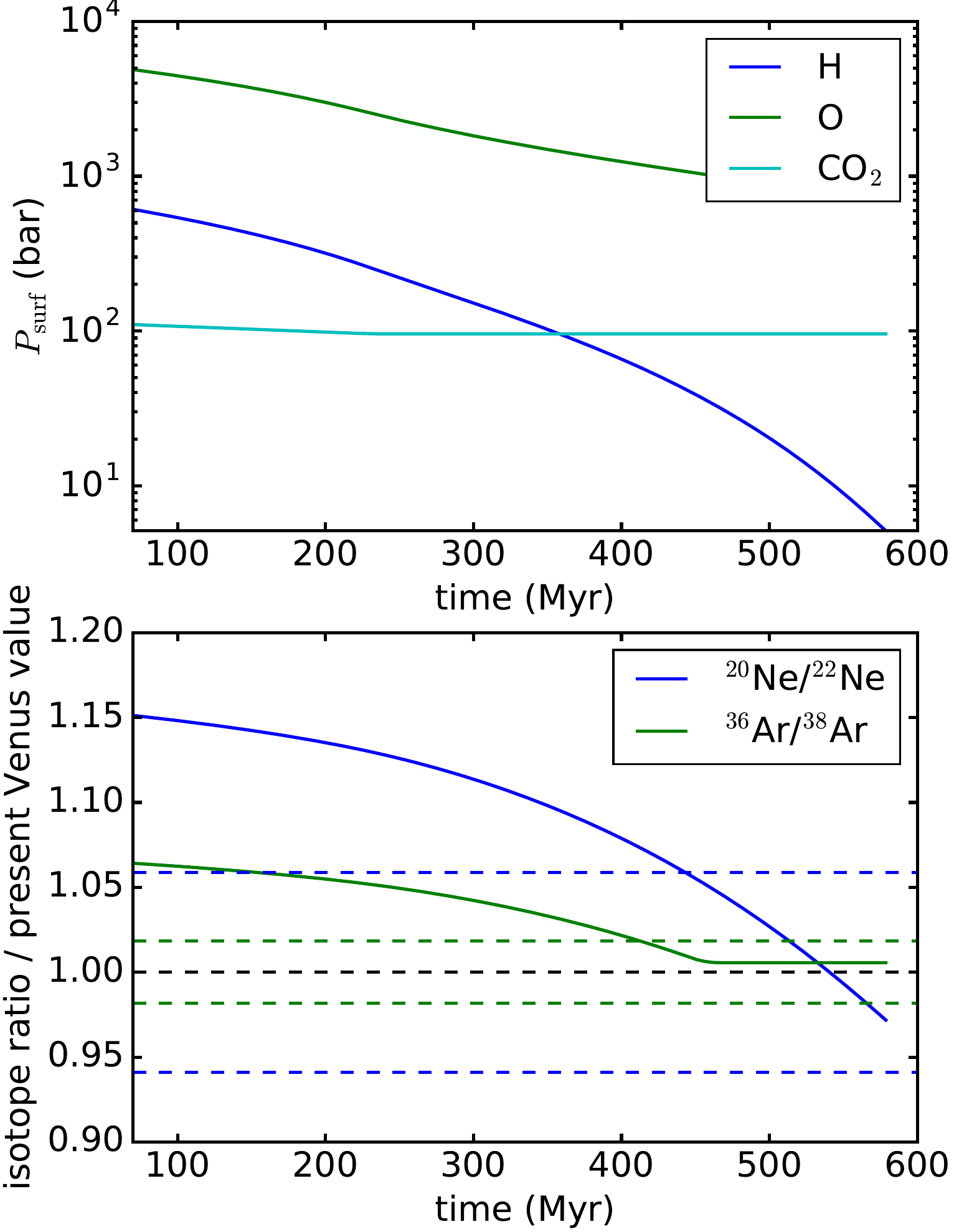}
	\includegraphics[height=9cm]{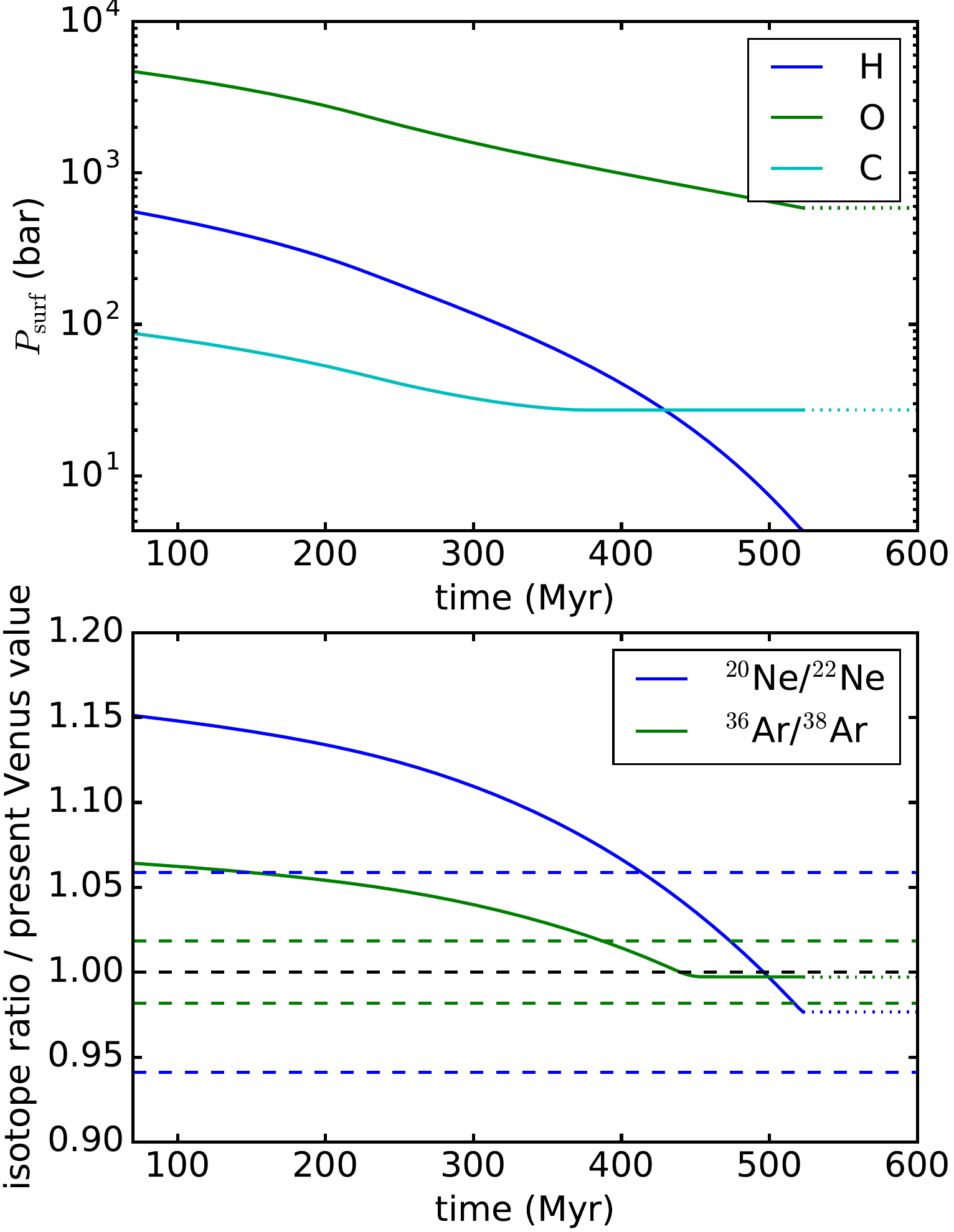}
	\caption{Isotope ratio evolution for a late accretion scenario around a fast rotator. Left: 5610~bar steam atmosphere with molecular CO$_2$. Right: 5320~bar steam atmosphere with dissociated CO$_2$.}
	\label{fig:venusisolate}
\end{figure}

The influence of assuming a higher temperature is not very strong, but it keeps the Ne ratio slightly higher and lowers Ar for otherwise similar conditions. Higher $T_0$ also requires a raised CO$_2$ content because it escapes more efficiently. Figure~\ref{fig:venusisolate} shows the example of a 5610~bar ($P_\mathrm{H_{2}O}=5500$~bar, $P_\mathrm{CO_{2}}=110$~bar) atmosphere with CO$_2$ molecules and a 5320~bar ($P_\mathrm{H_{2}O}=5000$~bar, $P_\mathrm{CO_{2}}=320$~bar) atmosphere with dissociated CO$_2$, both for a fast rotator. Such massive atmospheres are, however, very unlikely to be degassed from a Venus-mass planet \citep{Elkins-Tanton08}, especially in view of the likely drier large building blocks at Venus' orbit. Moreover, huge remnant O envelopes of about 1000~bar would also remain, if they were not completely oxidized in the hot surface. Therefore, present-day isotope ratios of Venus are unlikely to be caused only by escape of a steam atmosphere in a scenario where the planet accretes late.

\section{Discussion}
\label{sec:disc}

\subsection{Limitations and assumptions of the escape model}
The escape rates, and therefore atmospheric loss timescales, are influenced by our assumptions. Here we discuss them in more detail. Moreover, we test parameter variations for one reference case of a Mars-like embryo with a 96~bar atmosphere at the orbit of Earth for a moderate rotator.

\subsubsection{Heating and mass-loss efficiencies}
We adopt a heating efficiency of 15\% which is typical for hydrogen-dominated atmospheres \citep{Shematovich14}, but similar values were also used in other water loss studies \citep{Kasting83, Chassefiere96, Tian15a}. Raising $\eta$ by a factor of two to 30\%, like commonly adopted for O$_2$ \citep{Tian15}, raises the escape rates correspondingly and therefore shortens the loss timescale by the same value.

By assuming that the atmospheric escape is energy-limited (Eq.~\ref{eq:mdot}) we may also overestimate the escape rates if a non-negligible part of the energy provided by EUV absorption also increases the temperature and outflow velocity of the atmosphere \citep{Erkaev15}. \citet{Salz16} determined an evaporation efficiency factor, $\eta_\mathrm{eva}=\dot{M}_\mathrm{sim}/\dot{M}_\mathrm{el}$, the ratio of the mass-loss rate from hydrodynamic simulations to the energy-limited rate (cf. Eq.~\ref{eq:mdot}). From a study of early Mars \citep{Erkaev14} we find that for a H-dominated atmosphere at 100~EUV, $\eta_\mathrm{eva}{\sim}0.2{-}0.3$ with the lower value for $\eta=40$\% and the higher for 15\%. For O-dominated atmospheres (Appendix~\ref{sec:oescape}), however, $\eta_\mathrm{eva}{\sim}0.2{-}0.8$ and strongly depends on $\eta$ and $F_\mathrm{EUV}$, where larger $\eta_\mathrm{eva}$ is found for smaller values of these parameters. Due to the large scatter of these estimates we do not correct for this factor in our model. Introducing $\eta_\mathrm{eva}$ could therefore reduce the escape rates by up to a factor of five and lengthen the atmospheric loss timescales correspondingly.

\subsubsection{$r_0$ and $r_\mathrm{EUV}$}
By setting $\beta=r_\mathrm{EUV}/r_0=1$ we tend to underestimate the escape rates. \citet{Erkaev14} applied a hydrodynamic model to the upper atmosphere of an early Mars with an accreted hydrogen envelope. For pure H and an EUV flux enhanced by a factor 100 compared to the present solar value, they found that $\beta$ is 3--4 depending on the assumed $r_0$ and $\eta$. On the other hand, for pure O atmospheres (Appendix~\ref{sec:oescape}) we find that $\beta\approx 1.5$ and almost independent on $\eta$ or $F_\mathrm{EUV}$. Thus, the actual $\beta$ for a mixed atmosphere likely lies between these two extremes, indicating that we may underestimate the escape rates by about a factor of two. However, since $r_\mathrm{EUV}$ should also depend on the assumed wavelength of the EUV emission and corresponding photoionization cross-section in our gray model due to the wavelength-dependent penetration depth of the high-energy radiation, a more complex model would be required to determine $\beta$ more reliably. Additionally, the mass-loss rate depends on $r_0$, which should also change with time in an escaping atmosphere. If the Kelvin-Helmholtz contraction timescale is short compared to the mass-loss timescale, $r_0$ should decrease with time due to decreasing atmospheric mass, leading us to overestimate the escape rates at later evolutionary stages. On the other hand, if escape is very efficient and the atmosphere cannot adjust itself quickly enough, $r_0$ may also increase with time, leading to even more vigorous escape and the complete loss of the atmosphere. We test the influence of $r_0$ and $\beta$ for the reference case. The assumed mesopause height of 1000~km should be representative for the initial conditions of a steam atmosphere above a molten magma ocean after the models of \citet{Marcq12}, but could be overestimated at later stages when the atmosphere has already partly escaped and/or the surface temperatures have decreased. To test the effect we run the calculations for 500~km and 100~km. We find that the atmosphere is lost at 15 (17)~Myr for 500 (100)~km compared to 13.5~Myr for the reference case. On the other hand, if setting $\beta=2$ (intermediate between pure O and pure H atmospheres; Appendix~\ref{sec:oescape}) reduces the atmospheric loss timescale by a factor of four compared to the reference case.

\subsubsection{Mixing ratio in the escape region}
Atmospheric mixing ratios are calculated by assuming that the atmospheric species are well mixed in the escape region. This requires the assumption that the stellar EUV flux reaches close to the homopause level so that diffusive separation below the escape region is negligible. Since the increasing temperature in the outflow region increases the molecular diffusion coefficients, it is likely that the base of the flow and the homopause level are located at similar altitudes. If, however, diffusive separation may occur well below the EUV-heated escape region, the actual mixing ratios could be lower. The total escape rates would not be affected (if we ignore that $\eta$ may change), but the escape fluxes of the heavier constituents relative to H would be reduced. This means that whereas the O escape rate is 50\% that of H assuming $f_\mathrm{O}=0.5$, it would be 10\% that of H assuming $f_\mathrm{O}=0.1$ etc., since for our considered Mars-like objects $x_\mathrm{O}$ is always close to 1, i.e. drag is efficient. At a given total escape rate this would mean that H escapes more efficiently whereas less O is dragged with the flow if the actual mixing ratios are much lower than the maximum. Especially CO$_2$ could in such a case be present in the escape region in much smaller amounts due to its large mass, leading to less efficient removal. However, models of hot steam atmospheres indicate that such atmospheres are likely well mixed up to large heights \citep{Abe88, Kasting88, Marcq12}. \citet{Lupu14} studied the atmosphere of an Earth-like planet after a giant impact and found that it is dominated by H$_2$O and CO$_2$. Moreover, both constituents have almost constant mixing ratios up to large heights, if vertical mixing and photochemistry are taken into account. Therefore, it is likely that our assumptions about the mixing ratios at $r_0$ are valid for such atmospheres. However, a self-consistent treatment of the connection between the lower and upper atmosphere would be desirable and should be addressed in future studies.

According to \citet{Zahnle90}, the fractionation factor of the heavy major species (Eq.~\ref{eq:xj}) can also be written as $x_j=1-F_\mathrm{diff}/F_i$, where $F_\mathrm{diff}$ is the diffusion-limited escape flux of the light species. As long as $F_i>F_\mathrm{diff}$, the outflow of the light species is given by Eq.~\ref{eq:fi} and drags along species $j$. $F_i<F_\mathrm{diff}$ would yield $x_j<0$, i.e. $F_j=0$. In this case, the outflow of $i$ would not be energy-limited, but diffusion-limited and given by $F_\mathrm{diff}$, which is significantly lower. For Mars-like embryos, this situation never occurs in the early evolutionary stages, because of their low gravities and the high stellar EUV fluxes. Moreover, we show in Appendix~\ref{sec:oescape} that even an O-dominated atmosphere could escape hydrodynamically from such bodies. For proto-Venus, however, this situation may occur at later evolutionary stages. We also find that the negligible escape of CO$_2$ under most conditions leads to its accumulation. We do not continue the calculations if this situation occurs. However, since we are mainly interested to find the times at which the noble gas isotope ratios reach their present-day values (and remain constant thereafter), which occurs if both components of the ratio are not dragged anymore, we do not attempt to calculate the further evolution of the atmospheres. Since both Ne, Ar, and their considered isotopes are heavier than O, the latter still escapes with H during the relevant time period. Therefore, transition to diffusion-limited escape does not affect the conclusions from this study, but may become relevant if aiming to investigate the long-term evolution of Venus' early atmosphere.

\subsubsection{Atmospheric temperature}
We assume a temperature (entering the binary diffusion parameters) similar to the equilibrium temperatures at the respective orbits. This is consistent with the steam atmosphere model of \citet{Marcq12} and its updated version \citep{Massol16}, which predicts similar mesopause temperatures even above hot molten planetary surfaces. Adopting the temperature $T_0=T(r_0)$ for the calculations can be justified by inspecting Eq.~\ref{eq:esc}. Conservation of mass requires that the escape rate $Fr^2$ is constant with height, which means that the product $f_jx_j$ must also be constant with height. Since we set $f_j$ to $f_j(r_0)$ this requires setting $x_j=x_j(r_0)$ and thus $T=T_0$. However, the simple analytic description of the escape fluxes of the dragged species is derived assuming isothermal conditions and $df_j/dr=0$ \citep{Zahnle90}, so it could be partly an artifact of these simplifications. On the other hand, the situation could also be interpreted as using a mean $f_j$ (which is $\sim f_j(r_0)$ because it is roughly constant in an efficiently escaping upper atmosphere) and a mean $x_j$ (i.e., a mean $T$) of the upper atmosphere. The mean $T$ in the upper atmosphere would be higher because of heating by EUV absorption. Therefore, we tested the influence of this choice by using temperatures up to a few 1000~K, but find an negligible effect on the atmospheric loss timescales. The main effect is that CO$_2$ escapes more efficiently so that cases where CO$_2$ would accumulate are often circumvented and the evolution can be calculated up to higher ages. However, for atmospheres escaping efficiently already at low $T$, the timescales remain the same. Note that we use the present-day equilibrium temperatures which are slightly overestimated for the young ages considered here because the pre-main sequence Sun with an age $\geq$10~Myr had a lower luminosity than today. However, even a factor of two change in stellar luminosity would result in a change of $T_\mathrm{eq}$ by only about 50~K, which does not affect our results. The choice of temperature has a stronger impact on fractionation than on general atmospheric escape. Note that \citet{Gillmann09} could only reproduce the isotope fractionation patterns of Venus for restricted temperature ranges with their model.

\subsubsection{Escape process}\label{sec:disc_esc}
Our model may underestimate the mass-loss rates in case the escape is not mainly EUV-driven and thus cannot be described by Eq.~\ref{eq:mdot}. For young objects with extended atmospheres, ``boil-off'' driven by energy released from the gravitational contraction of the gas envelope and the stellar bolometric luminosity can yield escape rates orders of magnitude higher than those resulting from stellar EUV irradiation \citep{Owen16, Ginzburg16}. For low-gravity bodies, the atmospheres can also be removed by the rocky core's internal heat alone, independent of stellar irradiation, similar to the process removing accreted envelopes when the gas disk dissipates \citep{Stoekl15, Stoekl16}. A relevant parameter in this context is the Jeans escape parameter $\lambda=GM\mu m_\mathrm{H}/(kTr)$, where low values of $\lambda$ indicate efficient escape. Since this is especially crucial for hot low-gravity objects, we did not model planetary embryos smaller than Mars. Consider e.g. a Moon-like body with an outgassed steam atmosphere. At the body's surface, the escape parameter for H$_2$O is $\sim$20--30 for the equilibrium temperature range of 200--300~K. However, if the surface is still hot and molten, it would reduce to 1.5--4 for surface temperatures of 1500--4000~K \citep{Lebrun13}, comparable to the critical value where escape can become extremely high \citep{Levi09, Erkaev15}. The critical value for $\lambda$ in the case of a H$_2$O-dominated atmosphere is about 4 \citep[since $\gamma=1.33$ and $\lambda_\mathrm{crit}=\gamma/(\gamma-1)$; e.g.][]{Massol16} which indicates that water vapor is likely to be lost immediately after outgassing without building up an atmosphere, depending on the surface temperature of the magma ocean. The heavier CO$_2$ may build up an atmosphere, and as the surface cools, some build-up of steam may also be possible. However, due to the low gravity, also EUV-driven escape is much more efficient and these small atmospheres may be removed rather quickly. Direct escape of the steam atmosphere due to the hot molten surface is much less efficient for a more massive Mars-type body and irrelevant for full-grown planets like Earth.

\subsubsection{Start time}
We set a starting time of 10~Myr for our calculations under the assumption that the nebula gas has already disappeared from the considered orbital locations and a magma ocean exists and has outgassed its volatile inventory. It is, however, possible that these processes occur a few Myr earlier or later. Circumstellar disks typically disappear between 1--10~Myr \citep{Montmerle06, Hillenbrand08}, after which previously embedded embryos or protoplanets are exposed to the stellar EUV flux. Therefore, the effect of a different starting time lies mainly in lengthening/shortening the object's exposure to the saturated EUV emission (cf.~Section~\ref{sec:euv}), depending on the rotational evolution of the host star. In the saturation regime, the timescale for atmospheric loss remains the same because the EUV flux at a given orbit is roughly constant, only the system age at which the object loses its envelope shifts accordingly. Note that for the slow rotator evolution the saturation time is very short ($\sim6$~Myr) so that this evolution scenario is affected most.

\subsubsection{Orbits}
Planetary embryos likely have eccentric orbits due to dynamic interactions with other embryos, protoplanets, or gas giants \citep{Lunine11, Raymond14}. However, we assumed circular orbits with fixed semi-major axes for our calculations, yielding flux modulations only from stellar evolution. The orbit-averaged flux received by an object on an eccentric orbit scales as $(1-e^2)^{-1/2}$ \citep[e.g.][]{Dressing10} and is therefore larger than for a circular orbit, therefore we may underestimate the escape rates and overestimate the corresponding timescales for atmospheric loss. Moreover, non-zero eccentricity leads to tidal heating which may prolong the magma ocean phases \citep{Zahnle07}.

\subsubsection{Collisions}
In our calculations we assume that atmospheric escape takes place on an isolated object without any external disturbances (except for Section~\ref{sec:early} where we allow impacts to add mass to the core). However, in a young forming planetary system, collisions between embryos or with planetesimals and protoplanets are common, since this leads to the formation of larger terrestrial planets in timescales of 50--100~Myr \citep{Raymond14}. Collisions could either provide new volatile-rich material or remove parts of an existing steam atmosphere \citep{Genda03, Genda05}. Recent studies indicate that direct atmospheric erosion by an impact has a minor effect compared to the energy deposition leading to melting and subsequent degassing \citep{Gillmann16}. Moreover, large impacts which occur on timescales of 5--10~Myr \citep{Lebrun13} could lead to new magma ocean phases, triggering new episodes of outgassing. However, since the initial magma ocean degasses 70--99\% of its volatile inventory \citep{Elkins-Tanton08a}, later steam atmospheres are likely much more tenuous. Frequent collisions with small bodies could keep the surface molten and delay condensation of an outgassed steam atmosphere \citep{Lebrun13, Maindl15}. Therefore the condensation timescales indicated in Figs.~\ref{fig:mars1} and \ref{fig:mars2} are likely underestimated.

Note that for the proto-Venus evolution scenarios we also neglected impact-induced atmospheric loss. This can occur via the impact shock wave and/or enhanced hydrodynamic losses due to the increased temperatures from the released impact energy \citep[][and references therein]{Massol16}. In the former case, an Earth-like planet loses about 10\% of its atmosphere, whereas a Mars-like impactor loses about 30\%, so that roughly 70\% of the impactor's atmosphere is delivered to the planet \citep{Genda03, Genda05}. Moreover, impacts may (partially) melt the surface of a protoplanet, leading to a new magma ocean phase with outgassing from the interior. At Venus' orbit the magma ocean phase is expected to last for at least $10$~Myr \citep{Lebrun13} up to $\ge100$~Myr \citep{Hamano13}, larger than our assumed time interval between large impacts.

Atmospheric escape induced by impacts leads most likely only to loss without altering these patterns \citep{Canup00, Mikhail14}, but if the accreted embryos are already fractionated this may influence the final planet's patterns. Moreover, we assume that the Mars-like embryos degassed their volatile inventories and lost their complete atmospheres previously (this is likely valid at the orbit of Venus, cf.~Section~\ref{sec:embryos}). Then only the rocky material adds to the core and increases the protoplanet's gravity. Another possibility would be that Mars-like embryos could add an existing (possibly pre-fractionated) outgassed atmosphere upon impact. However, since the timescales for complete atmospheric loss from these objects is small compared to the typical time between large impacts on protoplanets (cf. Figs.~\ref{fig:mars1}, \ref{fig:mars2}), especially at the orbit of Venus, very massive outgassed atmospheres and/or very eccentric orbits which bring the embryo close to Venus' orbit only at perihelion would be required to supply a significant amount of atmosphere to the protoplanet which may alter its fractionation patterns.

\subsubsection{EUV evolution and wind of the young Sun}
In terms of EUV evolution, the present study considers the full range of possible conditions around the young Sun and solar-like stars in general. It is unknown if our Sun was once a fast or a slow rotator (or some case in between). Observations indicate a large spread (more than an order of magnitude) of possible activity levels between 10~Myr and 1~Gyr for a solar-mass star at a given age \citep{Tu15}. Hence, it is important to explore the full range of possible solar EUV histories, also for other Sun-like stars. The results of the embryo evolution, as well as the proto-Venus scenario, indicate a strong dependence on these assumptions. For the Mars-like embryo differences occur only between the slow and other activity levels because of the short atmospheric loss times which lie within the saturation period of the moderate and fast rotator tracks. The atmosphere is lost a few Myr earlier in the latter cases. Concerning fractionation of noble gases on Venus, all evolution tracks exhibit different results because the patterns can only be reproduced considering timescales much longer than the saturation time.

Previous studies on isotope fractionation or water loss on early Venus \citep{Chassefiere97, Gillmann09} assumed a dense stellar wind \citep[up to $10^3-10^4$ denser than now;][]{Chassefiere97} from the young Sun which contributed significantly to the modeled fractionation patterns due to additional energy input. However, more recent models of the winds of young Sun-like stars \citep{Johnstone15} suggests that they were much more tenuous than previously expected. The resulting energy input into planetary atmospheres by production of energetic neutral atoms is thus not very efficient and the stellar EUV luminosity is the dominant driver of atmospheric escape \citep{Lichtenegger16}.

\subsection{Water loss}
The short timescales found for the loss of water from catastrophically outgassed steam atmospheres on Mars-like embryos, especially for orbits smaller than Earth's and a moderate or fast rotator, suggests that planets formed from such (or even lower mass) bodies could have been drier than previously expected. The volatile amount these embryos are able to contribute to growing protoplanets will depend crucially on the timescales of planet formation, since the shorter the embryos are exposed to the stellar EUV emission the more volatiles can still be added to the protoplanets. Additionally, the timescales of outgassing and the amount of outgassed material (global vs. local magma oceans and their depths) will play a role because as long as sufficient volatiles remain inside the embryo these can be delivered to a protoplanet. Embryos smaller than Mars will be affected more severely because the atmospheric escape process is likely direct outflow due to the low gravity and (sufficiently) high surface temperature, similar to comets when they approach perihelion \citep[e.g.][and Section~\ref{sec:disc_esc}]{Hansen16}, and not EUV-driven mass-loss.

\subsection{Isotope ratios}
Thermal escape is one possible cause of isotope fractionation observed in present-day atmospheres of solar system objects. Previous studies already aimed to explain presently observed isotope and/or noble gas fractionation of the terrestrial planets by hydrodynamic escape \citep{Sekiya80, Hunten87, Sasaki88, Zahnle90, Pepin91, Gillmann09}. Using up-to-date stellar activity evolution tracks we explore, as an example, possible scenarios for the early evolution of Venus. We find cases for both a proto-Venus with impacts and a full Venus evolution where we are able to reproduce currently observed noble gas isotope ratios ($^{20}$Ne/$^{22}$Ne, $^{36}$Ar/$^{38}$Ar) if we assume that there is some residual H from the protoplanetary nebula mixed into an early outgassed steam atmosphere. Valid solutions are found for all stellar rotators and different assumptions about the planet's atmospheric radius. However, the ranges in allowed parameter space are always rather small. Moreover, the time at which the whole atmosphere is lost is often rather short ($<100$~Myr) after the present-day values are reached. If the atmosphere were completely lost, the current atmosphere and its fractionation pattern must result from other processes \citep[e.g.][]{Owen92, Halliday13}. Another possibility would be if the atmospheric escape mechanism would transition to the less-efficient Jeans escape at about these system ages. For instance, \citet{Kasting83} found that if the partial pressure of water vapor decreases to 10~bar, the escape rate drops significantly. Moreover, scenarios with too efficient loss of CO$_2$ are unrealistic, making the cases with slower stellar rotation and/or smaller atmospheric radii (corresponding also to smaller residual H envelopes) more likely. Escape of a pure steam atmosphere in a late accretion scenario also leads to valid solutions, but only for a fast rotator and in steam atmospheres in excess of a few kbar.

Despite our ability to reproduce present isotope ratios of Ne and Ar at similar times with our model, we are not able to reproduce the presently observed abundance ratio $^{20}$Ne/$^{36}$Ar of $0.5\pm0.3$ \citep{Hoffman79, Mikhail14} in any scenario, which remains almost solar-like \citep[$\sim$39;][]{Halliday13} in all cases. This ratio remains $\ge30$ in all scenarios although the isotopes of the individual noble gases reach the present-day values. Hydrodynamic escape is not capable of reproducing both abundance and isotope ratios at the same time if starting from solar values, in agreement with previous studies \citep{Zahnle90}. Whereas the isotope ratios ($^{20}$Ne/$^{22}$Ne, $^{36}$Ar/$^{38}$Ar) need to decrease by $\sim$10\%, the abundance ratio would have to decrease by a factor of $\sim$80 during similar timescales. This is clearly impossible considering the masses of the elements and the mass-dependent fractionation factors entering the escape fluxes in our model. A starting value of $^{20}\mathrm{Ne}/^{36}\mathrm{Ar}<1$ would be required. This suggests that the initial abundances were likely not solar, as suggested e.g. by \citet{Halliday13} who favor a mixing of chondritic and solar composition material. \citet{Pepin92} mention that cometary ices have low Ne/Ar, so accretion of such cometary material could be responsible for Venus' low Ne/Ar ratio. Since the evolution of the isotope ratios of a single element, however, is independent of the abundances, our results would not change if different element abundances at $t_0$ were assumed. Additional study of Kr or Xe isotopes would be valuable, though they are unknown for Venus. The abundance ratio $^{84}\mathrm{Kr}/^{130}\mathrm{Xe}$ is also not very well constrained for Venus, but a lower limit could be $\sim$100 \citep{Halliday13}. Our model results are consistent with this limit in most of the cases which reproduce Ne and Ar isotope ratios. Only for moderate and fast rotators with dissociated CO$_2$ does the Kr/Xe ratio become far too small. Also temperatures above 500~K would produce too low Kr/Xe ratios. The results of our model are in qualitative agreement with previous studies like \citet{Pepin91} who stressed that the Venusian noble gas isotope ratios, as well as abundances (except for Ne) are close to solar.

Aside from efficient thermal escape, other processes may lead to isotope fractionation. \citet{Genda05} modeled the atmospheric loss due to impacts in the presence of oceans. They found that if a water ocean is present on the surface, atmospheric losses due to impacts are enhanced. They argue that since it is likely that an ocean was present on Earth, but not on Venus, remnants of a noble gas-rich atmosphere may have survived on Venus, leading to different noble gas abundances on these planets. Although it is possible that liquid water was already present at Earth during the first $\sim$100~Myr, it is important to note that impacts of small embryos and planetesimals may have kept the planetary surfaces molten and the atmospheres in steam form. In this context, \citet{Debaille07} suggested, based on analysis of Sm-Nd isotopes, that Mars had a magma ocean phase of about $\sim$100~Myr, much longer than the timescales for water condensation estimated by \citet{Lebrun13} at this orbital distance ($\sim$0.1~Myr), who neglected such frequent impact heating events \citep{Maindl15}. Thus, it could be possible that Earth did not have oceans during the times of giant impacts yet, which would lead to an evolution scenario similar to Venus in the model of \citet{Genda05}. \citet{Genda05} assume that the planets formed by accumulation of several Mars-sized embryos, similar to our late-accretion scenario for Venus. Whereas they assume that these embryos supply the growing protoplanet with volatiles (aside from losses through the impacts), we found that such bodies lose their steam atmospheres very quickly at the orbit of Venus, and also at Earth's. This means that these planets could have accreted much drier than expected, if not sufficient bodies scattered from more distant orbits were involved, which would then result in rather tenuous outgassed atmospheres. However, recent Solar System formation models based on the ``Grand Tack'' hypothesis \citep{Walsh11, OBrien14} indicate that early Venus and Earth most likely grew to 50--70\% of their masses already within the protoplanetary nebula. This would have lead to accretion of gas envelopes of solar composition, which is in agreement with isotope studies at Earth \citep{Becker03}. These H$_2$-dominated envelopes would have to be removed by hydrodynamic escape until the surface pressures would allow solidification of the magma oceans and related outgassing of steam atmospheres. Large impacts may have occurred before or after a steam atmosphere was present, which would have affected the composition of the atmosphere in a different way. In reality, both processes may work together and determine the present atmospheric composition. Implementing impact-related losses in our model is planned for the future and will be used to study their effect on the evolution of isotope ratios.

A related issue in this context is the enrichment of deuterium relative to hydrogen. On Earth, D/H is comparable to chonditic material, but \citet{Genda08} showed that measured values can also be obtained by initially assuming mixtures of chondritic and solar material. The D/H ratio is then altered by evolutionary processes, such as surface exchange and atmospheric escape \citep{Genda08}. Both Mars and Venus show significant enrichments of D \citep[$\sim$6 for Mars; $\sim$150 for Venus;][]{Zahnle90} compared to Earth's value, so it is argued that evolutionary aspects must be responsible. \citet{Zahnle90} found that the Martian value can be reproduced with a wide range of initial values, as the outcome is very sensitive to the assumed parameters in their escape model. Since D is only marginally heavier than H, most studies argue that fractionation of D only occurs at later stages of evolution when H escape fluxes are low, because during periods of efficient hydrodynamic escape, D is also lost along with H, leading to negligible enrichment. \citet{Hunten93} stressed that for an enrichment of 150, like for Venus, the initial H content must have been at least 150 times higher than present, and even much higher if D escaped efficiently with H during the first tens to $\sim$100~Myr.

\section{Conclusions}\label{sec:conclusions}
We studied the evolution of outgassed H$_2$O/CO$_2$ steam atmospheres from planetary embryos due to EUV-driven hydrodynamic escape. Typical envelopes are lost in timescales of a few up to a few tens of Myr from Mars-like bodies, depending on the amount of degassed atmosphere, the orbital distance, and the detailed stellar activity evolution. For smaller Moon-mass embryos, atmospheres may be lost much quicker due to their lower gravity, or may not even accumulate at all. Depending on the typical timescales between collisions of large planetary embryos, protoplanets forming via late accretion may therefore be drier than expected. Our results indicate, in agreement with the recent findings of Ru isotopic data in carbonaceous chondrite-like asteroids by \citet{Fischer-Goedde17} and the ``Grand Tack'' hypothesis \citep[e.g.][]{OBrien14}, that the majority of volatiles should have been delivered from the outer Solar System during the main stages of terrestrial planet formation, but not during late accretion. Moreover, smaller planetesimals which did not experience magma oceans and fractionation also may have delivered volatiles that remained in their interiors and have not been outgassed before collisions. In this context, we investigate possible past evolution scenarios of Venus. Isotope ratios can be used to constrain possible histories of atmospheric escape. Under the assumption of late formation, present noble gas isotope ratios of Ne and Ar can only be reproduced if the Sun was a fast rotator and an outgassed atmosphere of a few kbar was present, which is unlikely. However, the present-day noble gas isotope ratios can also be reproduced in an early formation scenario with a nebula-accreted envelope for a variety of stellar EUV histories. The fast rotator case is the most unlikely because CO$_2$ would be lost too efficiently, inconsistent with the present 90~bar atmosphere. This amount is unlikely to be delivered later by large impacts because these objects should have lost their volatiles even more easily. It is more likely that the Sun's past evolution was in the range between a slow and moderate rotator. Also, it is unlikely that CO$_2$ was mainly dissociated since then it would also be lost too efficiently, further pointing towards a young Sun with a low activity level. A combined scenario of nebula-accreted material on a proto-Venus and magma ocean solidification and outgassing at a later time would also be an option, but this would depend strongly on the timings of the events. Measurements of Xe could confirm the early accretion scenario; if it were Earth-like then this would indicate that formation of a proto-Venus within the nebula was most likely. Therefore, in agreement with the ``Grand Tack'' hypothesis, early Venus most likely grew to a mass inside the nebula large enough such that it could capture a thin H$_2$-envelope, which was afterwards lost via EUV-driven hydrodynamic escape.

\section*{Acknowledgments}
PO and HL acknowledge support from the Austrian Science Fund (FWF): P27256-N27. NVE acknowledges RFBR grant No~16-52-14006~ANF\_a. AN and NT acknowledge support from the Helmholtz Association (project VH-NG-1017). This work was supported by the FWF NFN project S11601-N16 'Pathways to Habitability: From Disks to
Active Stars, Planets and Life' and the related subprojects S11604-N16 and S11607-N16. The authors also acknowledge the International Space Science Institute (ISSI, Bern, Switzerland) and the ISSI team 'The Early Evolution of the Atmospheres of Earth, Venus, and Mars'. We thank the two anonymous referees for helpful comments which significantly improved the paper.

\bibliography{mybibfile}

\appendix

\section{Hydrodynamic escape of oxygen from Mars-like embryos}
\label{sec:oescape}
Equations~\ref{eq:mdot} and \ref{eq:fi} are only valid when the main species in the upper atmosphere experiences efficient hydrodynamic escape so that the energy-limited equation approximates the mass-loss rates sufficiently well. For light species, such as atomic H, previous studies showed that Mars- and Earth-mass planets which are exposed to EUV fluxes enhanced by a factor of $>$25 compared to that of the present Earth will experience hydrodynamic expansion and escape conditions \citep[e.g.][]{Watson81, Zahnle86, Zahnle90, Chassefiere96a, Tian09, Erkaev13, Erkaev14, Luger15, Tian15, Tian15a}. For lower EUV fluxes, H$_2$ remains mainly in molecular form \citep{Yelle04, Tian05a, Koskinen10, Erkaev13}. Due to the high levels of stellar EUV emission during the first 50--100~Myr when embryos may still be present in a system, Eq.~\ref{eq:fi} is likely valid if hydrogen is the main species in the embryo's upper atmosphere.

However, in cases where escape is very efficient the amount of hydrogen decreases fast so that atomic oxygen accumulates and may become the main species. At least partly, some oxygen may be removed by oxidation of the surface \citep{Gillmann09}. However, for very high stellar EUV fluxes atomic oxygen may also escape hydrodynamically from low-gravity bodies like embryos and could potentially even drag along other heavy species, such as CO$_2$. To study this scenario we apply a 1-D hydrodynamic upper atmosphere model which takes into account heating by absorption of stellar EUV radiation. This model is described in detail in the appendix of \citet{Erkaev16}. We apply it to a Mars-like planetary embryo assuming an oxygen-dominated upper atmosphere to evaluate under which EUV flux levels such an atmosphere could escape hydrodynamically.

\begin{figure}
	\includegraphics[width=\textwidth]{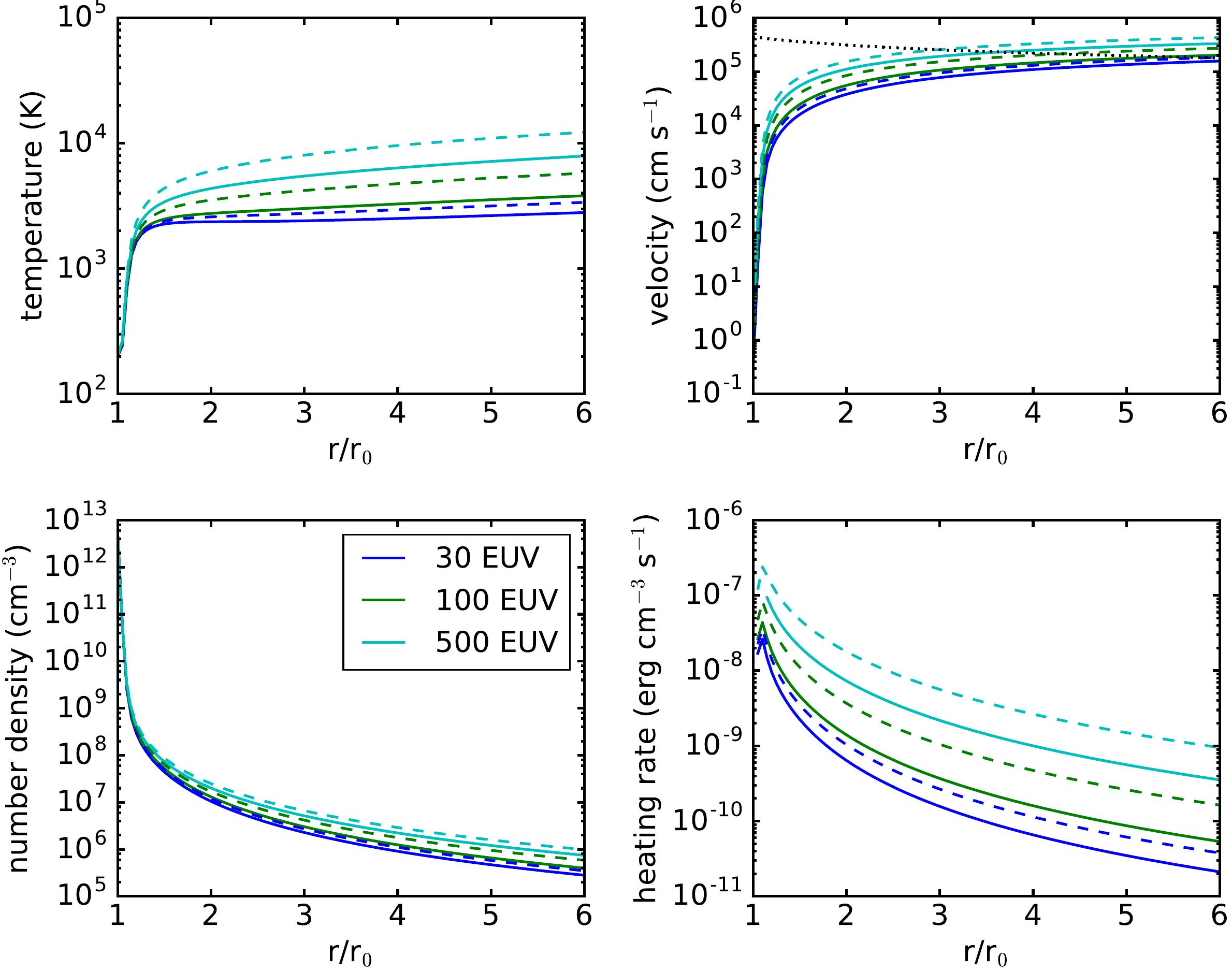}
	\caption{Atmospheric profiles of number density, velocity, temperature and volume heating rate of an O-dominated upper atmosphere for $\eta=15$\% (solid) and 30\% (dashed) and three EUV enhancement factors. They correspond to a Mars-like embryo located at the orbit of Mars. The black dotted line in the velocity plot corresponds to the escape velocity.}
	\label{fig:hydro}
\end{figure}

The model of \citet{Erkaev16} uses a wavelength-integrated EUV flux and does not consider the spectral dependence of the incoming stellar EUV radiation. We use an average atomic O photoabsorption cross-section of $10^{-17}$~cm$^2$ which is representative for the wavelength range of 150--900~\AA\ \citep{Fennelly92}. We set the lower boundary at the mesopause level at a radius of $r_0=4390$~km. Above this level, the bulk of the EUV photons is absorbed and very little penetrates below it. We assume an oxygen number density at the $r_0$ level of $5\times10^{12}$~cm$^{-3}$ \citep[typical homopause density; cf.][]{Erkaev13} and a temperature $T_0$ corresponding to the equilibrium temperature $T_\mathrm{eq}$ at the assumed orbital distance. In Fig.~\ref{fig:hydro} we show the results at the orbit of Mars where we have adopted $T_0=200$~K \citep{Marcq12}. The upper boundary of the calculation domain was set to $6~r_0$, where we assume outflow conditions. We present the results for two heating efficiencies, 15\% (as for H) and 30\% \citep[as for O$_2$;][]{Tian15}. The resulting atmospheric profiles of temperature, velocity, number density and volume heating rate are shown in Fig.~\ref{fig:hydro} for three different EUV flux levels (see below) and two heating efficiencies.

In Table~\ref{tab:lossrates}, we summarize the hydrodynamic modeling results for a Mars-like embryo at the orbit of Mars exposed to EUV fluxes from a young Sun of 50, 167, and 833~erg~s$^{-1}$~cm$^{-2}$ \citep[corresponding to flux levels approximately 30, 100 and 500 times higher than at present in the wavelength range 20--920~\AA][]{Ribas05}. Since the hydrodynamic equations are only valid for densities sufficiently high so that frequent collisions occur, the solutions are only considered to be valid if the outflow reaches the sonic point well below the exobase. We find that the sonic point lies above the exobase for case~Ia, so the escape rate should be more comparable to a (modified) Jeans rate and the upper atmosphere structure obtained with the hydrodynamic model is uncertain. For cases Ib, IIa, and IIb, the sonic point lies below the exobase, but the Knudsen number, which is the number where hydrodynamic and kinetic models start to deviate \citep{Volkov11}, is much higher than 0.1. Only for cases IIIa and b (the highest EUV flux level) the Knudsen number at the sonic point is sufficiently close to 0.1 so that these cases can be considered hydrodynamic blow-off. By comparing with the energy-limited escape rate we find that it approximates the hydrodynamically calculated rates better for low EUV fluxes and heating efficiencies, whereas it is up to a factor of five higher for high EUV fluxes. This finding is similar to \citet{Johnson13}. The results shown in Table~\ref{tab:lossrates} indicate that even heavy, O-dominated atmospheres can escape from Mars-like embryos if exposed to the high EUV fluxes of young stars because of their low gravity. Since the results shown are for Mars' orbit only, it is obvious that for closer orbits hydrodynamic escape of O would even be more efficient because of the even higher EUV flux levels. Therefore, it is justified to assume that outgassed steam atmospheres around Mars-like embryos at orbital locations spanning the HZ of the Sun are easily lost within a few to a few tens of Myr, regardless of their exact compositions and masses.

\begin{table}
	\caption{Hydrodynamic modeling results of an O-dominated upper atmosphere of a Mars-like embryo located at the orbit of Mars. For the six cases we varied the EUV enhancement factor and heating efficiency. The table includes the sonic point radius $r_\mathrm{s}$, the exobase distance $r_\mathrm{exo}$, the effective EUV absorption radius $r_\mathrm{eff}$, the hydrodynamic escape rate $L_\mathrm{hy}$ and the energy-limited escape rate $L_\mathrm{en}$ calculated with the same parameters.}\label{tab:lossrates}
	\centering
	\begin{tabular}{lcclllcc}
		\hline
		case & EUV & $\eta$ & $r_\mathrm{s}$ & $r_\mathrm{exo}$ & $r_\mathrm{eff}$ & $L_\mathrm{hy}$ & $L_\mathrm{en}$ \\
		&     & (\%) & $(r_0)$ & $(r_0)$ & $(r_0)$ & (s$^{-1}$) & (s$^{-1}$) \\
		\hline
		Ia   & 30  & 15 & 5.95 & 5.2  & 1.45 & $2.9\times10^{30}$ & $3.7\times10^{30}$ \\
		Ib   & 30  & 30 & 5.3  & 6.0  & 1.5  & $4.2\times10^{30}$ & $7.8\times10^{30}$ \\
		IIa  & 100 & 15 & 5.0  & 6.0  & 1.45 & $5.3\times10^{30}$ & $1.2\times10^{31}$ \\
		IIb  & 100 & 30 & 4.1  & $>$6 & 1.5  & $1.0\times10^{31}$ & $2.6\times10^{31}$ \\
		IIIa & 500 & 15 & 3.65 & $>$6 & 1.5  & $1.6\times10^{31}$ & $6.5\times10^{31}$ \\
		IIIb & 500 & 30 & 3.15 & $>$6 & 1.5  & $2.7\times10^{31}$ & $1.3\times10^{32}$ \\
		\hline
	\end{tabular}
\end{table}

\end{document}